\documentclass[reprint,prb,aps,amsmath,amssymb,floatfix,superscriptaddress,nofootinbib,longbibliography]{revtex4-2}

\usepackage{physics}
\usepackage{soul}
\usepackage{bm}
\usepackage{graphicx}
\usepackage{xcolor}
\usepackage[english]{babel}
\usepackage{amsfonts}
\usepackage{amssymb}
\usepackage[colorlinks = true,
            linkcolor = red,
            urlcolor  = blue,
            citecolor = magenta,
            anchorcolor = red]{hyperref}
\usepackage{amsmath}
\usepackage{siunitx}
\def\be{\begin{equation}}
\def\ee{\end{equation}}

\usepackage{braket}
\usepackage[normalem]{ulem}
\usepackage{siunitx}

\newcommand{\beq}{\begin{equation}}
\newcommand{\eeq}{\end{equation}}

\begin{document}

\title{Dephasing-powered open quantum-dot battery}

\author{Alok Nath Singh}
\affiliation{Department of Physics and Astronomy, University of Rochester, Rochester, NY 14627, USA}
\affiliation{Institute for Quantum Studies, Chapman University, Orange, CA 92866, USA}

\author{Rafael S\'anchez}
\affiliation{Departamento de F\'isica Teorica de la Materia Condensada, Universidad Aut\'onoma de Madrid, 28049 Madrid, Spain\looseness=-1}
\affiliation{Condensed Matter Physics Center (IFIMAC), Universidad Aut\'onoma de Madrid, 28049 Madrid, Spain}
\affiliation{Instituto Nicol\'as Cabrera (INC), Universidad Aut\'onoma de Madrid, 28049 Madrid, Spain }

\author{Andrew N. Jordan}
\affiliation{Institute for Quantum Studies, Chapman University, Orange, CA 92866, USA}
\affiliation{Department of Physics and Astronomy, University of Rochester, Rochester, NY 14627, USA}
\affiliation{The Kennedy Chair in Physics, Chapman University, Orange, CA 92866, USA}

\begin{abstract}
We propose a simple realization of an open quantum battery powered by local dephasing. The setup consists of a triple-quantum-dot chain coupled at one end to a thermal reservoir, with the dot at the opposite end strongly detuned to form a high-energy storage element. {Local dephasing populates this highly excited state and autonomously drives the system into a nonequilibrium steady state with high ergotropic efficiency. We find a limit where the ergotropic efficiency approaches unity, while the ergotropy itself approaches one-third of the energy detuning.} We characterize the battery through its energy-storage performance, charging time, and stability, showing that the charged state remains long-lived after the dephasing source is switched off. We further show that nearly all of the ergotropy can be extracted through a simple cyclic unitary protocol based only on modulation of the detuned-dot gate. Going beyond the unitary ergotropy bound, we then couple the detuned dot to an electronic load and show that about $90\%$ of the total stored energy can be converted into electrochemical work. Our results establish local dephasing as an autonomous resource for charging open quantum batteries and demonstrate simple routes for extracting the resulting stored energy in experimentally accessible mesoscopic platforms.
\end{abstract}

\begin{abstract}
We propose a simple realization of an open quantum battery powered by local dephasing. The setup consists of a triple-quantum-dot chain coupled at one end to a thermal reservoir, with the dot at the opposite end strongly detuned to form a high-energy storage element. Local dephasing populates this highly excited state and autonomously drives the system into a nonequilibrium steady state with high ergotropic efficiency. We find a limit where the ergotropic efficiency approaches unity, while the ergotropy itself approaches one-third of the energy detuning. We characterize the battery through its energy-storage performance, charging time, and stability, showing that the charged state remains long-lived after the dephasing source is switched off. We then demonstrate two complementary protocols for extracting the stored energy. First, nearly all the available ergotropy can be recovered through a simple cyclic unitary protocol involving only modulation of the detuned-dot gate. Second, coupling the detuned dot to an electronic load allows the extraction of electrochemical work exceeding the initial ergotropy, thereby accessing part of the battery's passive energy. Our results establish local dephasing as an autonomous resource for charging open quantum batteries and demonstrate simple routes for extracting the resulting stored energy in experimentally accessible mesoscopic platforms.
\end{abstract}

\maketitle


\section{Introduction}

A battery is a physical system that stores energy for later use. An effective battery stores a large amount of energy, retain it for sufficiently long times, and allow rapid charging and discharge. Quantum batteries seek to exploit quantum-mechanical effects to improve these figures of merit~\cite{campaioli_colloquium_2024}. A central motivation has been the possibility of genuine quantum advantages arising from collective dynamics, coherence, and entanglement. Examples include superextensive scaling of the charging power with system size and corresponding reductions in charging time under collective charging protocols~\cite{alicki2013entanglement,binder2015quantacell,campaioli2017enhancing,ferraro2018high}.

An important distinction in this context is between energy stored in a system and energy that can actually be extracted as work. The latter is quantified by the ergotropy, defined as the maximum work extractable from a quantum state through a cyclic unitary transformation~\cite{allahverdyan2004maximal}. A wide variety of quantum-battery architectures and charging mechanisms has consequently been explored, including coherent driving~\cite{ferraro2018high}, interaction-mediated collective charging~\cite{andolina2019extractable}, dissipative charging~\cite{rossini2019many}, thermalization-assisted protocols~\cite{hovhannisyan2020charging}, and reservoir-engineered open quantum batteries~\cite{xu2021enhancing}. In open quantum batteries, the environment itself participates in charging and stabilization, creating opportunities for autonomous operation but also introducing dissipation and passive energy that reduce the useful fraction of the stored energy~\cite{santos2019stable,gherardini2020stabilizing,barra2019dissipative}.

\begin{figure}[b]
    \includegraphics[width=\linewidth]{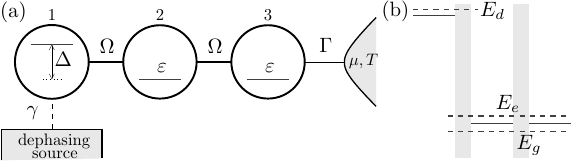}
\caption{
Dephasing-powered open quantum battery consisting of a triple-quantum-dot chain. (a) The rightmost dot is coupled to an electron reservoir at temperature $T$ and chemical potential $\mu$ with a tunneling strength $\Gamma$. The leftmost dot is detuned by $\Delta$, while the other two have a common energy $\varepsilon$. A nearby dephasing source is coupled to it with strength $\gamma$ and serves as the sole charging resource. (b) Schematic representation of the quantum dot levels (solid lines) and the hybridized eigenenergies $E_{\rm g}$, $E_{\rm e}$, and $E_{\rm d}$ (dashed lines).
}
    \label{fig:setup}
\end{figure}

A particularly interesting source of nonequilibrium dynamics is dephasing. Although usually regarded as detrimental for suppressing quantum coherence, it can also redistribute population between energy eigenstates whenever the dephasing operator does not commute with the system Hamiltonian~\cite{singh2026capturing, singh2026collective}. 
Accordingly, there has been a growing interest in dephasing- and noise-assisted quantum batteries. 
Depending on the physical platform, environmental noise has been shown to accelerate charging, stabilize stored energy, improve robustness against disorder or parameter detuning, and even induce nontrivial steady-state charging dynamics~\cite{mayo2022collective,shastri2025dephasing,
ahmadi2025harnessing,liu2025charging,fajar2025noise}. More broadly, dissipative and nonequilibrium environments have emerged as useful energetic resources for quantum batteries, including common reservoirs, heat currents, and steady-state transport~\cite{tabesh2020environment,tacchino2020charging,kamin2023ness}. Dephasing can arise naturally from uncontrolled environmental fluctuations. Continuous quantum measurement represents a more engineered setting that can generate the same unconditional dephasing dynamics when the measurement record is disregarded. Measurement backaction nevertheless provides a useful thermodynamic precedent, since it can supply energy to a quantum system and power thermodynamic processes~\cite{elouard2017extracting,buffoni2019quantum,yi2017single,ferreira_transport_2024,elouard_revealing_2026,sanchez2026making,litzba_coupling_2026}.

In this work, we propose an open quantum battery in which energy is stored in the occupation of a highly excited electronic state. The battery, sketched in Fig.~\ref{fig:setup}, consists of a triple quantum dot (TQD) chain in the strong-Coulomb-blockade regime, such that at most one electron occupies the array~\cite{gaudreau_stability_2006,kouwenhoven2001few}. The leftmost quantum dot (QD1) is strongly detuned from the remaining dots and therefore provides the energy-storage element: charging the battery amounts to transferring population into the high-energy state predominantly localized on this dot. Local dephasing of QD1 supplies the energy required for this transfer by redistributing population among the TQD energy eigenstates. A thermal reservoir coupled to the opposite end of the array exchanges electrons efficiently with the low-energy states while coupling only weakly to the highly detuned state. The combination of dephasing-induced population transfer and state-selective reservoir coupling thereby produces the population imbalance required for a non-passive charged state. Our goal is to populate this otherwise weakly occupied high-energy state rapidly, stabilize its occupation after charging, and provide a direct channel through which its stored energy can subsequently be discharged on demand.

Our results demonstrate that local dephasing can generate highly active nonequilibrium steady states, with an ergotropic efficiency approaching 1 in the low-temperature limit of the thermal reservoir. We further show that the charged state remains long-lived after dephasing is switched off. We demonstrate two complementary routes for extracting the stored energy. A simple cyclic modulation of the detuned-dot gate enables the extraction of nearly all the stored ergotropy. A complementary approach involves coupling the detuned dot to an electronic load, through which electrochemical work exceeding the initial ergotropy can be obtained by also drawing on the battery’s passive energy. Altogether, these findings establish local dephasing as an autonomous thermodynamic resource for charging open quantum batteries and provide an experimentally accessible platform based on mesoscopic quantum transport architectures.

The remainder of the manuscript is organized as follows. We present the model in Sec.~\ref{sec:model} and the charging dynamics in Sec.~\ref{sec:performance}, including the charging time and stability. The discharge of the battery into an electronic load and the resulting electrochemical work extraction are analyzed in Sec.~\ref{sec:discharge}, followed by conclusions in Sec.~\ref{sec:conclusion}.

\section{Model}
\label{sec:model}

The system of interest consists of a linear triple quantum dot (TQD), as shown in Fig.~\ref{fig:setup}. We consider nearest-neighbor hopping with amplitude $\Omega$. Counting the quantum dots from left to right, as we do throughout the paper, QDs 2 and 3 have the same onsite energy $\varepsilon$, whereas QD1 is detuned upward by an energy $\Delta\gg\Omega$ and acts as the energy-storage site. This configuration can be tuned through voltages applied to nearby plunger gates~\cite{ihn_semiconductor_2009}. In the strong Coulomb blockade with up to one electron, the spin degree of freedom is irrelevant and will not be considered. In the local basis, the TQD Hamiltonian is
\begin{equation}
\hat H_0=\sum_{j=1}^{3}(\varepsilon+\Delta\delta_{j1})\hat d_j^\dagger \hat d_j - \sum_{j=1}^{2} \Omega \left(\hat d_j^\dagger\hat d_{j+1} +\mathrm{h.c.}\right),
\end{equation}
where $\hat d_j^\dagger$ creates an electron on dot $j$ and $\delta$ denotes the Kronecker delta.

QD3 is tunnel-coupled to a thermal reservoir, while the remaining dots are isolated from it. The reservoir is characterized by temperature $T$ and chemical potential $\mu$, with Hamiltonian $\hat H_{\mathrm{res}}=\sum_k\varepsilon_k\hat c_k^\dagger\hat c_k$. The tunnel coupling is described by $\hat H_{\mathrm{tun}}=\sum_k\zeta\hat d_3^\dagger\hat c_k+\mathrm{h.c.}$, where $\hat c_k^\dagger$ creates an electron in reservoir mode $k$. In the weak-coupling regime, this interaction defines the bare tunneling energy scale $\Gamma=2\pi|\zeta|^2\nu$, where $\nu$ is the reservoir density of states, assumed to be constant in the wide-band limit approximation. The corresponding physical tunneling rate is $\Gamma/\hbar$. Throughout the article, unless otherwise stated, we set the reservoir chemical potential to $\mu=\varepsilon$, aligning it with the onsite energies of QDs 2 and 3.

As shown in Fig.~\ref{fig:setup}, we subject the detuned QD1 to local dephasing with strength $\gamma$. The dephasing source acts as the nonequilibrium external resource that supplies energy to the battery during charging, whereas the thermal reservoir supplies electrons. In the main text, we leave its microscopic origin unspecified and characterize it only through the dephasing strength $\gamma$. Two possible realizations are discussed in the appendices. One is a noisy resistor, for which Johnson–Nyquist voltage fluctuations generate local charge dephasing where the dephasing strength is proportional to the temperature and resistance of the resistor (see Appendix~\ref{app:noise}). The other is continuous charge measurement using a quantum point contact~\cite{ihn_quantum_2009}, where the biased detector generates measurement backaction and the local dephasing strength is proportional to the charge fluctuations on QPC, or the signal-to-noise ratio for an efficient detector (see Appendix~\ref{app:QPC}). Although these implementations have different microscopic interpretations, they produce the same ensemble-averaged dephasing term considered here.

\subsection{Eigenstates in the High-Detuning Regime}

The exact eigenstates can be expressed in the local basis as $|a\rangle=u_a|1\rangle+v_a|2\rangle+w_a|3\rangle$, although the coefficients have cumbersome expressions in terms of the system parameters.
We are interested in the limit $\Delta\gg\Omega$, where the eigenstates of $\hat H_{0}$, up to first order in $\Omega/\Delta$, are 
\begin{equation}\label{eq:eigenstates}
\begin{aligned}
|g\rangle=&\left(1+\frac{\Omega^2}{2\Delta^2}\right)^{-1/2}
\left(\frac{|2\rangle+|3\rangle}{\sqrt2}
+\frac{\Omega}{\sqrt2\Delta}|1\rangle\right),\\
|e\rangle=&\left(1+\frac{\Omega^2}{2\Delta^2}\right)^{-1/2}
\left(\frac{|2\rangle-|3\rangle}{\sqrt2}
+\frac{\Omega}{\sqrt2\Delta}|1\rangle\right),\\
|d\rangle=&\left(1+\frac{\Omega^2}{\Delta^2}\right)^{-1/2}
\left(|1\rangle-\frac{\Omega}{\Delta}|2\rangle\right).
\end{aligned}
\end{equation}
The state $\ket{d}$ is the highly detuned eigenstate and is predominantly localized on QD1. The other two eigenstates, $\ket{g}$ and $\ket{e}$, are primarily symmetric and antisymmetric combinations of QDs 2 and 3, respectively, and therefore have substantial overlap with the reservoir-coupled QD3. The high-energy state $\ket{d}$ has only a small admixture of the low-energy dots, of order $\Omega/\Delta$, which strongly suppresses its coupling to the reservoir. 


Taking the ground-state energy as the reference, $E_g=0$, fixes the common onsite energy $\varepsilon$ accordingly. In the high-detuning limit, the remaining eigenenergies are separated into a low-energy state $\ket{e}$ at an energy of order $2\Omega$ and a highly excited state $\ket{d}$ at an energy of order $\Delta$, cf. Fig.~\ref{fig:setup}(b) where the energy levels $E_{\rm g}$, $E_{\rm e}$, and $E_{\rm d}$ are indicated with dashed lines. To leading order in $\Omega/\Delta$, the large energy separation of $\ket{d}$ from the low-energy subspace is therefore controlled directly by the gate detuning $\Delta$. Note that our choice of the chemical potential, $\mu=\varepsilon$, places it approximately midway between $E_{\rm g}$ and $E_{\rm e}$.

The large separation of $\ket{d}$ from the two low-energy states makes its occupation energetically unfavorable under ordinary coherent dynamics. Its hybridization with the low-energy subspace is suppressed by $\Omega/\Delta$, and its overlap with the reservoir-coupled QD3 is of order $\mathcal O(\Omega^2/\Delta^2)$. This separation is central to the battery operation: local dephasing can populate the high-energy state during charging, whereas its weak reservoir coupling {(mediated by higher-order tunneling processes involving the two intermediate dots)} protects the stored excitation once the dephasing source is removed.

\subsection{Passive energy and ergotropy}
\label{sec:ergotropy}

The useful energy stored in the battery is quantified by its ergotropy, defined relative to the corresponding passive state. We first compute the total internal energy
\begin{equation}
E(t)=\mathrm{Tr}\!\left[\hat H_{0}\hat\rho(t)\right],
\label{eq:energy}
\end{equation}
where the reduced density matrix of the TQD, $\hat\rho(t)$, describes the evolution of the system state.

The passive state is obtained by arranging the eigenvalues of $\hat{\rho}(t)$ in decreasing order over the energy eigenstates of $\hat H_{0}$ in increasing order of energy. Its energy is therefore
\begin{equation}
E_{\rm passive}(t)
=
\sum_n r_n(t) E_n,
\label{eq:passiveE}
\end{equation}
where $r_n(t)$ are the eigenvalues of $\hat{\rho}(t)$ sorted in decreasing order and $E_n$ are the eigenenergies of $\hat H_{0}$ sorted in increasing order. The ergotropy is then~\cite{allahverdyan2004maximal}
\begin{equation}
\mathcal W(t)
=
E(t)-E_{\rm passive}(t),
\label{eq:ergotropy}
\end{equation}
which is nonnegative by construction. A state is active, and can therefore
serve as a source of extractable work under cyclic unitary operations, when
$\mathcal W(t)>0$.

\subsection{Open-System Dynamics}
\label{sec:master_equation}

We describe the ensemble-averaged dynamics using a global Lindblad master equation formulated in the eigenbasis of the coupled-dot Hamiltonian. This treatment incorporates interdot hybridization into the reservoir-induced transition rates and provides a thermodynamically consistent weak-coupling description when the coherent coupling between the dots cannot be neglected~\cite{levy2014local,potts:2021,rivas:book,breuer:book}. The global master equation is valid in the weak system–reservoir coupling regime, where the reservoir-induced transition rates are small compared to the relevant energy-level splittings, and under the usual Born–Markov and secular approximations.

The reduced density matrix evolves according to~\cite{schaller:book}
\begin{equation}
\frac{d\hat\rho}{dt}
=
-\frac{i}{\hbar}[\hat H_0,\hat\rho]
+
\frac{1}{\hbar}\left[
\sum_a
\left(
\mathcal D[\hat L_{a}^{\mathrm{in}}]
+
\mathcal D[\hat L_{a}^{\mathrm{out}}]
\right)
+
\mathcal D[\hat L_\gamma]\right]\hat\rho,
\label{eq:global_master_equation}
\end{equation}
where $\mathcal D[\hat O]\hat\rho=\hat O\hat\rho\hat O^\dagger-\{\hat O^\dagger\hat O,\hat\rho\}/2$ and $a\in\{g,e,d\}$.
Since the reservoir is coupled only to QD3, particle injection and extraction are described by the jump operators
\begin{equation}
\hat L_{a}^{\mathrm{in}}
=
\sqrt{W_{a0}}\,\ket{a}\bra{0},
\qquad
\hat L_{a}^{\mathrm{out}}
=
\sqrt{W_{0a}}\,\ket{0}\bra{a},
\label{eq:reservoir_jump_operators}
\end{equation}
where the tunneling strengths involving eigenstate $\ket{a}$, 
\begin{equation}
W_{a0}
=
\Gamma |w_a|^2 f(E_a),
\qquad
W_{0a}
=
\Gamma |w_a|^2[1-f(E_a)],
\label{eq:charging_rates}
\end{equation}
where $w_a=\braket{3|a}$ denotes the overlap of eigenstate $\ket{a}$ with the local state of the reservoir-coupled QD3~\cite{brandes2005coherent}. Here 
$f(E)=[1+\exp((E-\mu)/k_{\mathrm B}T)]^{-1}$ is the Fermi--Dirac distribution describing the thermal state of the reservoir with temperature $T$ and electrochemical potential $\mu$.  

The reservoir selectivity follows directly from the overlaps $w_a$. The states $\ket{g}$ and $\ket{e}$ are predominantly supported on QDs 2 and 3 and therefore both have substantial overlap with the reservoir-coupled QD3. By contrast, the highly detuned state $\ket{d}$ is predominantly localized on QD1, and its amplitude on QD3 is of order $\mathcal O(\Omega^2/\Delta^2)$, and hence does not appear at the order given in Eq.~(2). Population transferred into $\ket{d}$ therefore exchanges particles with the reservoir much more slowly than population in the low-energy states, allowing energy deposited in the detuned state to remain stored in the TQD.

Local dephasing strongly modifies this virtual-transport picture by coupling $\ket{d}$ to the low-energy eigenstates. 
The dephasing operator, generated by the barrier-dot occupation operator, $\hat n_1=\hat d_1^\dagger\hat d_1$, is, in the energy eigenbasis:
\begin{equation}
\hat L_\gamma
=
\sqrt{\gamma}\hat n_1
=
\sqrt{\gamma}
\sum_{a,b}
{u_a^*u_b^{} }
\ket{a}\bra{b},
\label{eq:dephasing_jump_operator}
\end{equation}
where $u_a=\langle 1|a\rangle$ and $u_b=\langle 1|b\rangle$ denote the overlaps of the energy eigenstates with the local state of QD1, cf.~Eq.~\ref{eq:eigenstates}. Although $\hat n_1$ is diagonal in the local occupation basis, it is nondiagonal in the energy eigenbasis because $[\hat n_1,\hat H_0]\neq0$. The dephasing channel therefore conserves particle number while redistributing populations among the single-electron energy eigenstates. In particular, the off-diagonal components of $\hat n_1$ couple the predominantly localized high-energy state $\ket{d}$ to the low-energy states $\ket{g}$ and $\ket{e}$. Local dephasing can consequently transfer population from the reservoir-accessible low-energy subspace into the highly excited state without directly injecting an additional electron into the TQD. Note that $\Gamma$ and $\gamma$ are expressed in energy units, so that the corresponding physical rates are obtained by dividing by $\hbar$ in Eq.~\ref{eq:global_master_equation}.

In previous works, the same basic mechanism was shown to enhance the occupation of an otherwise weakly occupied detuned state and to modify transport through triple-quantum-dot systems~\cite{singh2026capturing,singh2026collective,sanchez2026making}. The present configuration uses this effect for a different purpose. Rather than placing the detuned dot between two transport sites, we place the highly excited state at one end of the TQD and couple a single thermal reservoir to the opposite end. The resulting spatial separation makes the low-energy states strongly reservoir accessible while the detuned state remains only weakly coupled to the reservoir. Dephasing can therefore populate the high-energy state faster than the reservoir can empty it, producing the nonequilibrium population ordering required for a non-passive steady state.

\section{Dephasing-Induced Charging and Battery Performance} \label{sec:performance}

We now investigate the charging properties of the dephasing-powered quantum battery. Unless stated otherwise, we consider a strongly detuned storage dot with $\Delta=100\Omega$, reservoir coupling $\Gamma=0.1\Omega$, reservoir temperature $k_{\rm B}T=\Omega$, and chemical potential aligned with the low-energy onsite level, $\mu=\varepsilon$. Local dephasing acts on QD1 with strength $\gamma=10\Omega$, while the remaining dots are not directly dephased. These parameters place the system deeply in the high-detuning regime, where the excited state localized on QD1 is only weakly hybridized with the reservoir-accessible low-energy states. The configuration in which dephasing acts only on QD1 is the simplest realization of the charging mechanism: it acts directly on the dot supporting the high-energy storage state and efficiently redistributes population between this state and the low-energy manifold. Dephasing of the other individual dots is compared in Appendix~\ref{app:deph_config}.

We employ a single-orbital approximation which is also consistent with typical GaAs/AlGaAs energy scales. In these devices, the interdot hopping can be tuned to the $\mu{\rm eV}$ range~\cite{Mukhopadhyay2018APL2x2,Studenikin2021SST}, and bath coupling can be made smaller, of order $0.1\,\mu{\rm eV}$~\cite{Nakajima2017PRL}. At the same time, the first orbital excitation is typically much larger, in the meV range~\cite{Elzerman2004PhysicaE,WillemsVanBeveren2005NJP}. Thus, a physical scale $\Omega\sim1\,\mu{\rm eV}$ gives $\Gamma=0.1\Omega\sim0.1\,\mu{\rm eV}$ and $\Delta=100\Omega\sim0.1\,{\rm meV}$, while the first excited orbital remains more than an order-of-magnitude above the detuned state. Therefore, higher local orbitals are not expected to participate in the dynamics considered here.

We also checked that the charging dynamics remain confined predominantly to the empty and single-electron sectors. Although dephasing can redistribute population within a fixed charge subspace, it does not itself change the total particle number. Population of higher-charge sectors is instead controlled by particle injection from the reservoir and by the corresponding charging-energy cost. Using a spinful extended-Hubbard model including all two-electron states, we find that the two-electron occupation remains negligible over the reservoir window $\mu/\Omega\in[-0.5,0.5]$ and $k_{\rm B}T/\Omega\in[0.1,5]$ (see Appendix~\ref{app:twoelectron}).

\begin{figure}
    \centering
    \includegraphics[width=\linewidth]{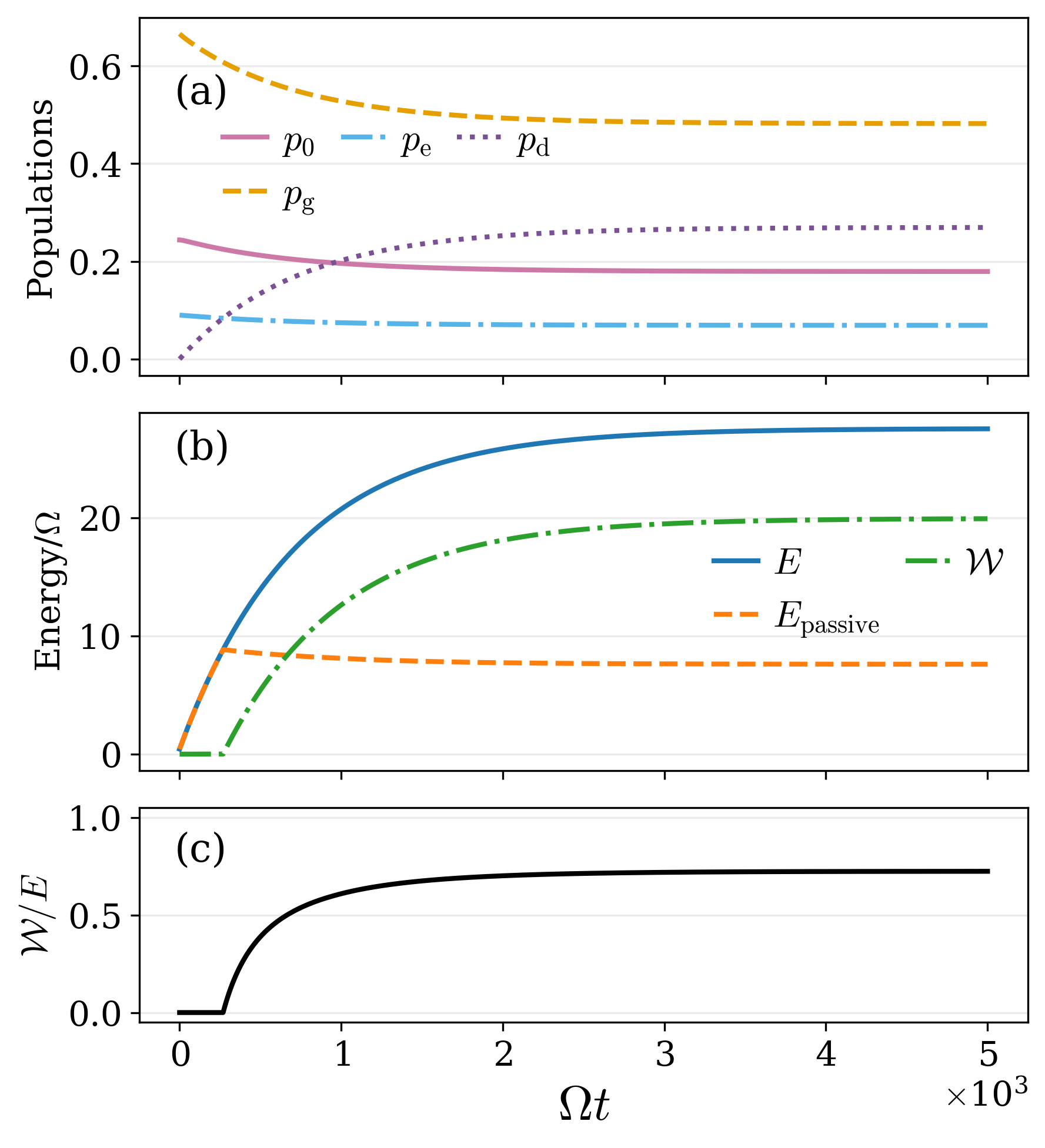}
    \caption{
Charging dynamics of the dephasing-powered quantum battery for
$\Delta=100\Omega$, $\Gamma=0.1\Omega$, $k_{\rm B}T=\Omega$,
$\mu=\varepsilon$, and $\gamma=10\Omega$.
(a) Populations of the empty state and the three single-electron energy
eigenstates. 
(b) Internal energy $E=\mathrm{Tr}(\hat H_0\hat{\rho})$, passive-state energy
$E_{\rm passive}$, and ergotropy $\mathcal W$. Dephasing drives the system
into a highly excited nonequilibrium steady state while maintaining a
large extractable-work component.
(c) Ergotropic efficiency $\mathcal W/E$, which approaches $\approx$
$73\%$ in the steady state.
}
    \label{fig:ergotropy}
\end{figure}

Figure~\ref{fig:ergotropy} summarizes the charging dynamics. The evolution is obtained from the global master equation presented in Sec.~\ref{sec:master_equation}. Starting from thermal equilibrium with the reservoir, dephasing drives the system away from equilibrium and steadily increases its internal energy.
Figure~\ref{fig:ergotropy}(a) shows the populations of the empty state and the three single-electron energy eigenstates of $\hat H_0$. 
Dephasing transfers substantial population into the high-energy detuned state $\ket{d}$, while the reservoir continues to exchange particles efficiently with the low-energy states $\ket{g}$ and $\ket{e}$. With $\mu=\varepsilon$ lying approximately midway between these two states, the reservoir preferentially populates the lower-energy bonding state $\ket{g}$, leaving the antibonding state $\ket{e}$ relatively weakly occupied. Together with the substantial occupation of $\ket{d}$ induced by dephasing, this population distribution produces a highly non-passive steady state.
%

The internal energy increases monotonically before saturating to a nonequilibrium steady-state value substantially above the equilibrium energy, see Fig.~\ref{fig:ergotropy}(b). Importantly, most of the injected energy remains available as extractable work rather than being converted into passive energy.

The delayed onset of ergotropy marks the time at which the state first becomes non-passive. At early times, dephasing raises the internal energy, but the population distribution remains passive and therefore contains no extractable work. 
The ergotropic efficiency $\mathcal W/E$ is shown in Fig.~\ref{fig:ergotropy}(c). After an initial transient, the efficiency rapidly increases and approaches a steady-state value exceeding $70\%$. This result is remarkable because open-system batteries coupled to thermal reservoirs typically suffer from a large passive-energy contribution that significantly reduces the extractable fraction of the stored energy. 

Once the battery has been charged, the dephasing source can be switched off, as the charged state remains protected by the large energy detuning. Specifically, the decay rates of the eigenstates \eqref{eq:eigenstates} into the reservoir, proportional to $|w_a|^2\Gamma$, are strongly suppressed for the charged state, {resulting in $W_{0d}=\Gamma{\cal O}(\Omega^4/\Delta^4)$}. The long-term stability of the stored energy is examined in Sec.~\ref{sec:stability}.

For the parameters considered here, the steady-state battery energy reaches approximately $E_{\rm ss}\simeq27.76\Omega$, of which about $\mathcal W_{\rm ss}\simeq20.17\Omega$ remains extractable, corresponding to an ergotropic efficiency of approximately $0.73$. The large ergotropy is primarily caused by the dephasing-induced population of the highly detuned state $\ket{d}$. In the steady state, its population reaches $p_d\simeq0.27$ despite lying roughly $\Delta=100\Omega$ above the low-energy subspace, whereas the remaining populations are approximately $p_0\simeq0.18$, $p_g\simeq0.48$, and $p_e\simeq0.07$. The resulting occupation of the highest-energy state is therefore far larger than would occur in thermal equilibrium and produces a strongly non-passive population distribution.

It is instructive to compare the present mechanism with our earlier works on measurement-assisted tunneling in triple-quantum-dot architectures~\cite{singh2026capturing,sanchez2026making,singh2026collective}. In those systems, the strongly detuned state was associated with the central dot and measurement-induced dephasing converted its otherwise virtual occupation into real population, thereby modifying transport through the array. Here we use a different geometry and operating principle: the detuned state is localized on the end dot QD1, while the only thermal reservoir is attached to the opposite end, QD3. The low-energy states therefore remain strongly coupled to the reservoir, whereas the high-energy storage state is nearly isolated from it. This state-selective coupling allows dephasing-induced excitation of $\ket{d}$ to accumulate as a persistent population imbalance and produces a non-passive steady state with large ergotropy.


\begin{figure}
    \centering
    \includegraphics[width=\linewidth]{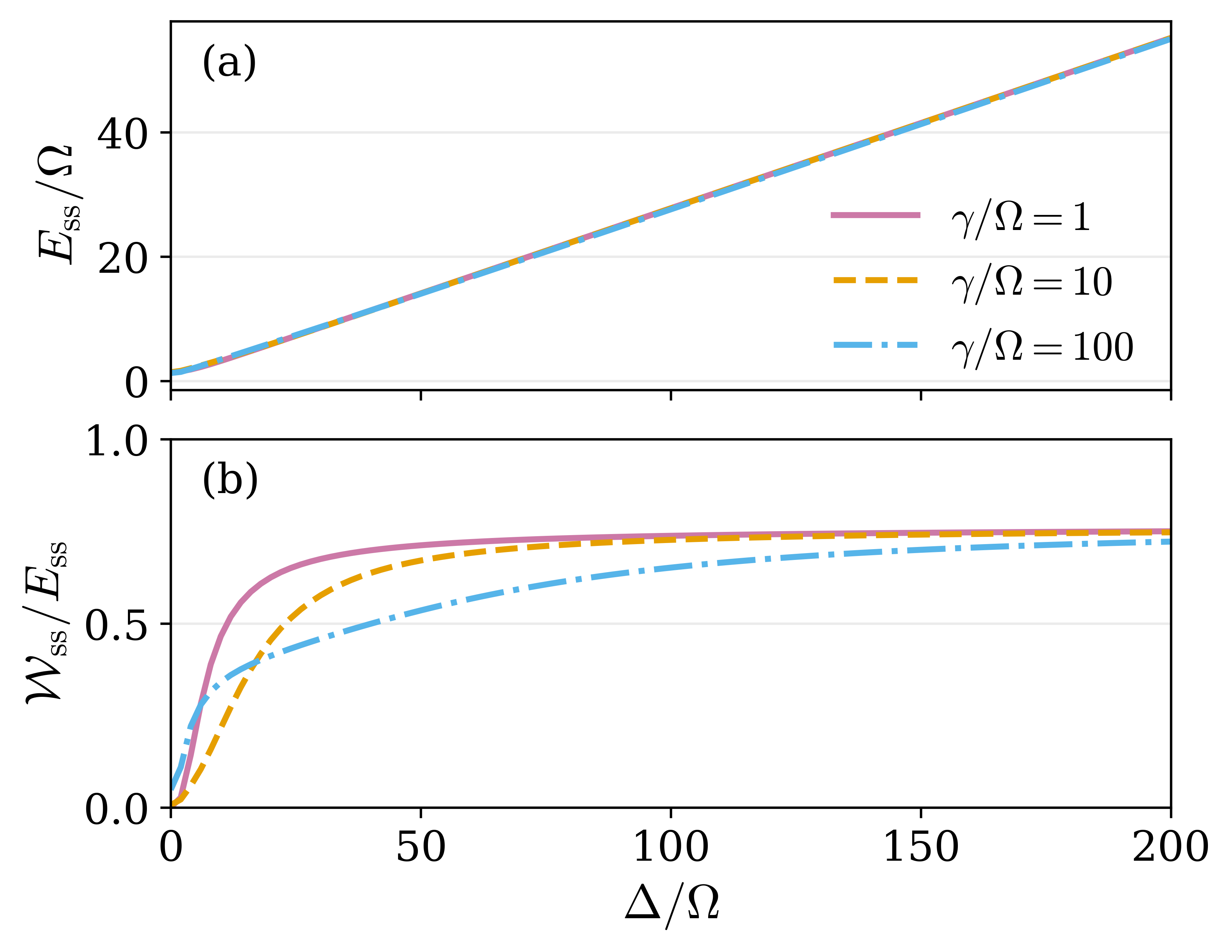}
    \caption{
    Steady-state battery performance as a function of barrier detuning $\Delta$, for different cases of dephasing strength $\gamma$. The remaining parameters are fixed at $\Gamma=0.1\Omega$, $k_{\rm B}T=\Omega$, and $\mu=\varepsilon$, with dephasing acting on the barrier dot. Panel (a): steady-state internal energy $E_{\rm ss}/\Omega$. The stored energy increases mainly with $\Delta$, reflecting the increasing energy of the populated barrier-like state, while its dependence on $\gamma$ is much weaker. Panel (b): ergotropic efficiency $\mathcal W_{\rm ss}/E_{\rm ss}$. The efficiency remains high over a broad parameter range, indicating that the dephasing-induced active state is robust against changes in $\Delta$ and $\gamma$.
}
\label{fig:steady_state_line}
\end{figure}

\subsection{Steady-State Performance}
\label{sec:steady_state}

We now examine the steady-state battery performance over a broader parameter range. Figure~\ref{fig:steady_state_line} shows the steady-state internal energy $E_{\rm ss}$ and ergotropic efficiency $\mathcal W_{\rm ss}/E_{\rm ss}$ as a function of the barrier detuning $\Delta$, for different dephasing strengths $\gamma$, with all other parameters kept fixed at $\Gamma=0.1\Omega$, $k_{\rm B}T=\Omega$, and $\mu=\varepsilon$.
The steady-state energy is controlled primarily by the barrier detuning. Increasing $\Delta$ raises the energy of the barrier-like state and therefore increases the energy stored when this state is populated, as seen in Fig.~\ref{fig:steady_state_line}(a).
 The ergotropy increases linearly with $\Delta$ for $\Delta\gtrsim100\,\Omega$, since the internal energy increases linearly in Fig.~\ref{fig:steady_state_line}(a) and ergotropic efficiency becomes constant in Fig.~\ref{fig:steady_state_line}(b). 
This linear scaling is expected to hold only while $\Delta$ remains
well below the orbital excitation energies of QD2 and QD3, beyond
which the single-orbital approximation may no longer be valid.

By contrast, the dependence on $\gamma$ is weak over the range shown: changing the dephasing strength modifies the steady-state energy negligibly compared to changing $\Delta$. The ergotropic efficiency shows a similar behavior, see Fig.~\ref{fig:steady_state_line}(b). It remains large over a broad region of parameter space and varies only weakly with $\gamma$. Thus, the high extractable fraction is not a fine-tuned feature; it lies within a broad plateau where the steady state remains strongly non-passive.


In the large-detuning regime, the steady-state energy and ergotropy have compact analytic expressions to leading order in $\Omega/\Delta$.
Unlike the numerical calculations, we leave the reservoir temperature $T$ arbitrary in this analysis. For chemical potentials between the two low-energy eigenstates, $E_g<\mu<E_e$, the steady-state energy and ergotropy to leading order in $\Omega/\Delta$ are
\begin{equation}
\begin{aligned}
E_{\rm ss}&\simeq
\frac{e^{(\mu-\varepsilon)/k_{\rm B}T}
\cosh\left(\frac{\Omega}{k_{\rm B}T}\right)}
{1+3e^{(\mu-\varepsilon)/k_{\rm B}T}
\cosh\left(\frac{\Omega}{k_{\rm B}T}\right)}
\Delta,\\
\mathcal W_{\rm ss}&\simeq
\frac{e^{(\mu-\varepsilon)/k_{\rm B}T}
\sinh\left(\frac{\Omega}{k_{\rm B}T}\right)}
{1+3e^{(\mu-\varepsilon)/k_{\rm B}T}
\cosh\left(\frac{\Omega}{k_{\rm B}T}\right)}
\Delta.
\end{aligned}
\end{equation}
The corresponding ergotropic efficiency approaches
\begin{equation}
\frac{\mathcal W_{\rm ss}}{E_{\rm ss}}
\simeq\tanh\left(\frac{\Omega}{k_{\rm B}T}\right).
\end{equation}
For $\mu=\varepsilon$ and $k_{\rm B}T=\Omega$, this gives an ergotropic efficiency of approximately $0.76$, close to the numerical value of $0.73$ obtained in Fig.~\ref{fig:ergotropy}. These expressions explain the nearly linear increase of the stored energy and ergotropy with $\Delta$, as well as the saturation of the ergotropic efficiency at large detuning. Interestingly, the leading-order ergotropic efficiency is independent of the reservoir chemical potential $\mu$, even though the stored
energy depends on it. In the limit where $k_{\rm B}T\ll\Omega$, these expressions predict that the ergotropic efficiency approaches unity, while the stored energy approaches $\Delta/3$.

Although the final steady-state energy and efficiency are only weakly affected by $\gamma$, the charging time of the battery is strongly controlled by the dephasing strength, which we examine below.

\subsection{Charging Time}
\label{sec:charging_time}

\begin{figure}
    \centering
    \includegraphics[width=.9\linewidth]{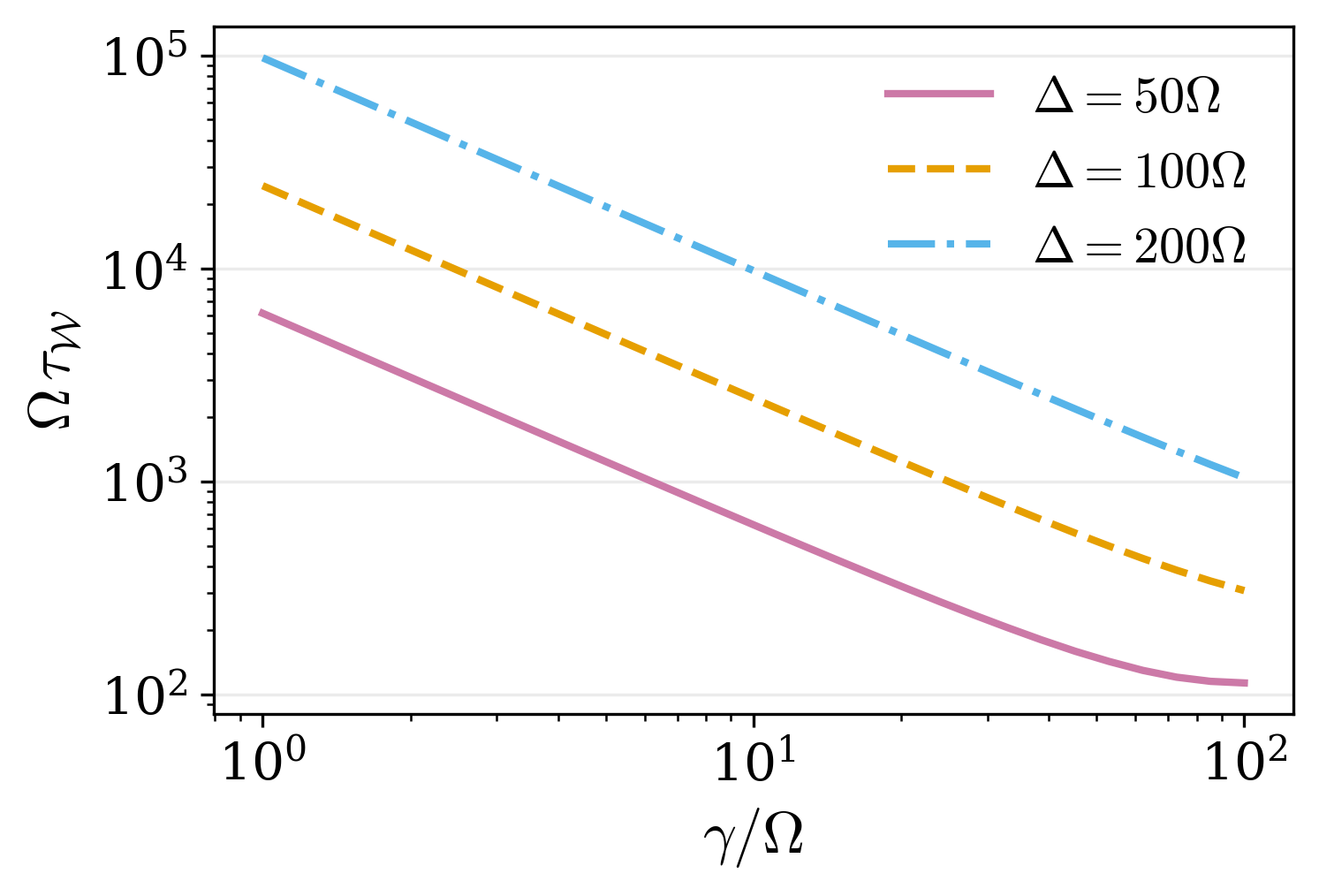}
    \caption{
Ergotropic charging time as a function of dephasing strength for several barrier detunings. Increasing the dephasing strength substantially reduces the charging time, showing that $\gamma$ provides a kinetic control knob for the battery. Larger detuning slows the charging process because the high-energy barrier state is more weakly hybridized with the low-energy states. Parameters: $\Gamma=0.1\Omega$, $k_{\rm B}T=\Omega$, and $\mu=\varepsilon$.
}
    \label{fig:charging_time}
\end{figure}

To see whether $\gamma$ affects the charging rate, we define an ergotropic charging time
\begin{equation}
\tau_{\mathcal W}
=
\min\left\{
t:
\frac{\mathcal W(t)}
{\mathcal W_{\rm ss}}
\ge 0.95
\right\},
\label{eq:charging_time}
\end{equation}
where the initial thermal state has zero ergotropy. This definition focuses on the time required to generate useful extractable work, rather than merely the time required to increase the internal energy.
Figure~\ref{fig:charging_time} shows $\tau_{\mathcal W}$ as a function of the dephasing strength for several barrier detunings. The charging time decreases approximately as a power law with increasing $\gamma$, showing that dephasing controls the rate at which the active nonequilibrium state is produced. Larger detuning leads to longer charging times because the detuned state is more weakly hybridized with the low-energy states. As a result, dephasing-induced population transfer into the detuned state becomes slower. Across the detunings shown, increasing $\gamma$ consistently shortens $\tau_{\mathcal W}$, while increasing $\Delta$ shifts the curves upward.

In the large-detuning regime, the dephasing-induced transition rates into the storage state scale as $\gamma\Omega^2/\Delta^2$. Since these transitions control the slow charging dynamics, charging time scales as $\tau_{\mathcal W}\propto\hbar\Delta^2/(\gamma\Omega^2)$, explaining the approximately inverse dependence on the dephasing strength observed in Fig.~\ref{fig:charging_time}.

The dephasing strength is an experimentally tunable parameter. In the noisy-resistor realization discussed in Appendix~\ref{app:noise}, increasing $\gamma$ corresponds to increasing the voltage fluctuations seen by the dots, for example by increasing the effective noise temperature, resistance, or capacitive lever arm. In the continuous-measurement realization discussed in Appendix~\ref{app:QPC}, increasing $\gamma$ corresponds to increasing the measurement strength, which can be achieved by increasing the detector response or the QPC bias, within the regime where the Markovian dephasing description remains valid.

\subsection{Stability of the Charged Battery}
\label{sec:stability}

A practically useful quantum battery must retain its stored energy after the charging process has been terminated. To assess the stability of the charged state, we investigate the relaxation dynamics following the removal of the dephasing source. The battery is first charged until its ergotropy reaches $95\%$ of the steady-state value. The dephasing is then switched off ($\gamma=0$), while the Hamiltonian, reservoir coupling, chemical potentials, and temperatures remain unchanged. The subsequent dynamics therefore describe the natural relaxation of the charged battery in the absence of dephasing.

\begin{figure}
    \centering
    \includegraphics[width=\linewidth]{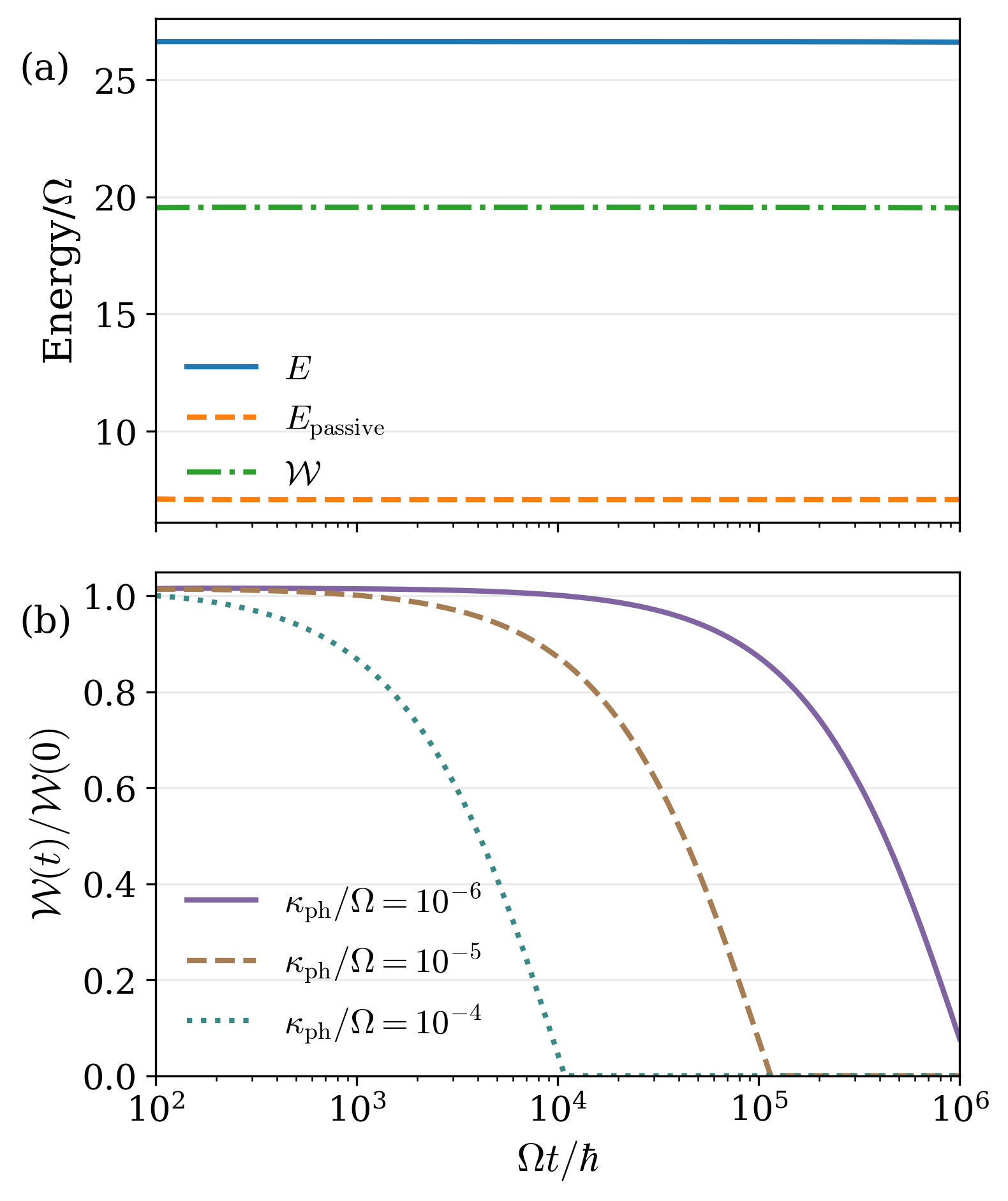}
    \caption{
Stability of the charged quantum battery after the dephasing source is switched off.
(a) Internal energy $E$, passive energy $E_{\rm passive}$, and ergotropy $\mathcal W$ in the absence of phonon relaxation, demonstrating strongly suppressed relaxation through the thermal reservoir.
(b) Ergotropy retention in the presence of phonon-induced relaxation for different values of $\kappa_{\rm ph}$.
In each case, phonon relaxation is included during both charging and storage, and the battery is initially charged to $95\%$ of its corresponding steady-state ergotropy.
Other parameters are the same as in Fig.~\ref{fig:ergotropy}.
}
    \label{fig:stability_phonons}
\end{figure}

Figure~\ref{fig:stability_phonons}(a) shows the evolution of the internal energy, passive energy, and ergotropy after the dephasing source is removed. Over the time window shown, the battery remains stable in the charged state: the internal energy, ergotropy, and passive energy remain essentially unchanged. This stability arises because the large detuning $\Delta\gg\Omega$ localizes the high-energy eigenstate $\ket{d}$ predominantly on QD1, while the thermal reservoir is coupled only to the distant QD3. The overlap of $\ket{d}$ with the reservoir contact is consequently $\mathcal O(\Omega^2/\Delta^2)$, strongly inhibiting relaxation of the stored excitation through the reservoir. Thus, dephasing can be switched off during storage without appreciable loss of the stored work over the interval considered here.

{For the representative scale $\Omega=1\mu{\rm eV}$, the reservoir temperature $k_{\rm B}T=\Omega$ corresponds to approximately $11.6{\rm mK}$, while the upper limit $\Omega t/\hbar=10^6$ corresponds to approximately $0.66{\rm ms}$. The negligible decay observed over this interval shows that relaxation through the thermal electronic reservoir is not expected to limit the battery lifetime on this timescale.

A common source of relaxation in semiconductor quantum dots is electron–phonon coupling, which can induce transitions between the energy eigenstates. To examine its effect on storage stability, we introduce phenomenological phonon-induced transitions from the high-energy state $\ket{d}$ to the low-energy states $\ket{g}$ and $\ket{e}$, characterized by a total spontaneous emission rate $\kappa_{\rm ph}/\hbar$ (see Appendix~\ref{app:phonon_relaxation}). Figure~\ref{fig:stability_phonons}(b) shows the resulting ergotropy retention for $\kappa_{\rm ph}/\Omega=10^{-6},10^{-5}$, and $10^{-4}$. We have verified numerically that including phonon relaxation even at $\kappa_{\rm ph}=10^{-4}\Omega$ produces only small corrections to the steady-state ergotropy and internal energy. For example, the ergotropic efficiency decreases by $3.6\%$ for the parameters in Fig.~\ref{fig:ergotropy}. Thus, phonon relaxation is neglected elsewhere in the manuscript.

Experiments on GaAs double quantum dots have reported strongly suppressed phonon-assisted tunneling, with dimensionless relaxation strength $\kappa_{\rm ph}/\Omega$ as low as approximately $3\times10^{-6}~$\cite{hofmann2020phonon}. Although this motivates the range considered here, the experimental device operates with substantially smaller interdot hopping, and the effective phonon relaxation rate depends on the device geometry and electronic spectrum. The achievable rate in the present TQD architecture therefore remains device-dependent. As shown in Fig.~\ref{fig:stability_phonons}(b), the storage lifetime decreases approximately inversely with $\kappa_{\rm ph}$, while substantial ergotropy can still be retained over timescales much longer than the characteristic tunneling time.}

\section{Discharging the Quantum Battery}
\label{sec:discharge}

We next examine how the energy stored by dephasing-induced charging can be
released as useful work. We consider two complementary discharge protocols.
First, we study cyclic unitary extraction, for which the ergotropy provides
the fundamental upper bound on the extractable work. Remarkably, approaching
this bound does not require a complicated optimal control sequence: nearly
all of the stored ergotropy can be extracted through a simple and
intuitive protocol involving only the plunger gate of the detuned dot. The
gate is rapidly lowered to bring the dot near resonance, the system is allowed
to evolve coherently, and the gate is then raised back to its initial value,
completing a cyclic transformation of the Hamiltonian. We then consider an open-system protocol in which QD1 is coupled
to an electronic load. In this case, particle exchange with the load allows
the stored energy to be converted directly into electrochemical work. Unlike
cyclic unitary extraction, this process is not bounded by the initial
ergotropy and can also access part of the passive energy of the battery. 

\subsection{Cyclic Unitary Discharge}
\label{sec:unitary_discharge}

The maximum work extractable from the charged state through a cyclic unitary
transformation is given by its ergotropy. The corresponding optimal
transformation maps the charged state onto its passive state while returning
the Hamiltonian to its initial form, reducing the battery energy by exactly
\begin{equation}
W_{\rm ext}^{\rm max}
=
\mathcal W(0)
=
E(0)-E_{\rm passive}(0).
\end{equation}
Although this provides a natural benchmark for discharge, implementing the
general optimal unitary may require independent control of several
transitions. We therefore ask how much of the ergotropy can instead be
accessed through a simple control already available in a quantum-dot device:
modulation of the plunger gate controlling the detuned QD1.

\begin{figure}
    \centering
    \includegraphics[width=\linewidth]{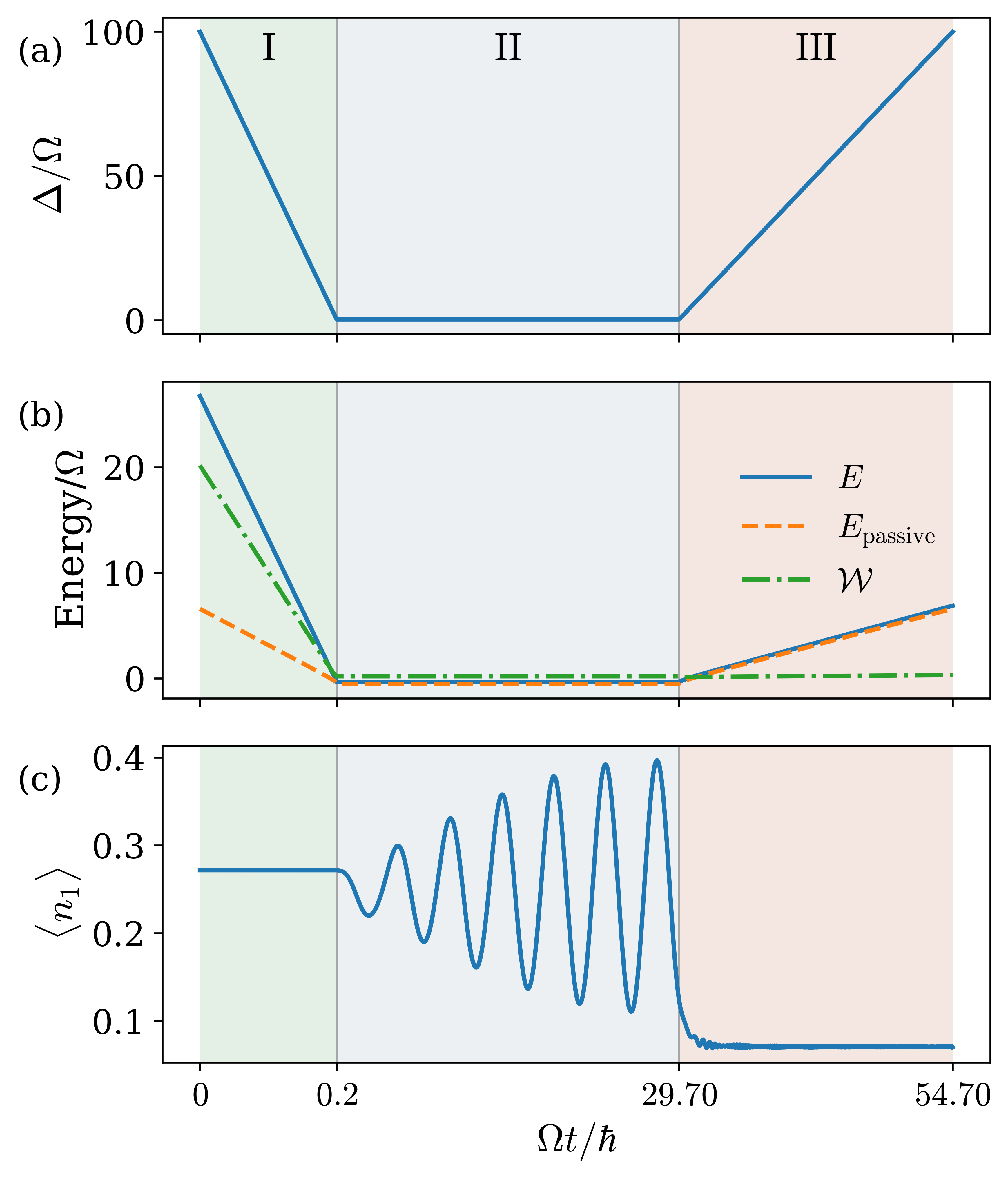}
    \caption{
Cyclic unitary discharge by modulation of the QD1 gate voltage.
(a) Detuning protocol, divided into three stages: (I) rapid reduction of
$\Delta$ from $100\Omega$ to $\Delta_{\min}=0.23\Omega$,
(II) coherent evolution at fixed $\Delta_{\min}$, and
(III) restoration of the initial detuning.
(b) Internal energy $E$, passive energy $E_{\rm passive}$, and ergotropy
$\mathcal W$ during the protocol.
(c) Occupation of the detuned dot, $\langle n_1\rangle$.
The parameters are
$\Omega\tau_{\rm down}/\hbar=0.2$,
$\Omega\tau_{\rm hold}/\hbar=29.5$, and
$\Omega\tau_{\rm up}/\hbar=25$.
For visibility, the horizontal scale is independently rescaled within the
three stages; time is linear within each stage, the tick labels give the
physical values of $\Omega t/\hbar$, and there is no time gap between the stages.
}
    \label{fig:unitary_discharge}
\end{figure}

We consider a cyclic protocol in which the energy of the storage dot
is lowered while it remains highly occupied, allowing work to be
extracted from the battery. The system is then allowed to evolve
coherently at a lower detuning, redistributing population away from
the storage dot. Finally, its energy is raised back to the original
value. If the occupation of the storage dot is sufficiently reduced
before the return, less work is required to restore its energy than
was extracted during the initial reduction, resulting in positive
net work over the cycle.

To implement this protocol, we first disconnect the dephasing source
and the reservoir, so that the battery evolves unitarily. We then
control only the detuning $\Delta(t)$ of QD1, keeping all other
Hamiltonian parameters fixed. Since the only time-dependent term
in the Hamiltonian is $\Delta(t)\hat n_1$, the instantaneous power
supplied to the battery is
\begin{equation}
\dot W_{\rm sys}(t)
=
{\rm Tr}\!\left[\hat\rho(t)\dot{\hat H}(t)\right]
=
\dot\Delta(t)\langle\hat n_1(t)\rangle.
\label{eq:unitary_gate_power}
\end{equation}
The net work extracted over the complete cycle is therefore
\begin{equation}
W_{\rm ext}
=
-\int_0^{t_{\rm f}}
dt\,\dot\Delta(t)\langle\hat n_1(t)\rangle
=
E(0)-E(t_{\rm f}),
\label{eq:unitary_extracted_work}
\end{equation}
where $\hat H(t_{\rm f})=\hat H(0)$. {Physically, the extracted work is associated with the electrostatic energy change induced by the plunger gate. Using $\dot\Delta(t)=-e\alpha\dot V_g(t)$, where $V_g$ is the gate voltage and $\alpha$ is the capacitive lever arm defined in Appendix~\ref{app:noise}, the work can equivalently be expressed as $W_{\rm ext}=e\alpha\int_0^{t_{\rm f}}dt,\dot V_g(t)\langle\hat n_1(t)\rangle$.}

A realization of this approach is shown in Fig.~\ref{fig:unitary_discharge} using a three-stage gate protocol. Starting from the charged steady
state, the detuning is rapidly lowered in stage I, held near resonance
during stage II to allow coherent population redistribution, and
finally restored to its initial value in stage III. During the rapid downward ramp, the occupation of QD1 changes only weakly, allowing substantial work extraction. At the reduced detuning, coherent tunneling produces oscillations in the QD1 occupation, and the hold time is chosen near one of its minima. During the upward ramp, the occupation remains substantially below its initial value, so only a fraction of the extracted work is supplied back to the battery.

For the protocol shown in Fig.~\ref{fig:unitary_discharge}, the charged state
has initial energy $E(0)\simeq27.79\Omega$ and ergotropy
$\mathcal W(0)\simeq20.50\Omega$. The gate cycle extracts
$W_{\rm ext}\simeq20.20\Omega$, corresponding to approximately $98.6\%$ of
the initial ergotropy. The parameters above
were selected from a numerical search for a simple high-efficiency protocol,
rather than from a strict global optimization. During the cycle, $\mathcal W$ can exceed $E$ because the fixed energy reference is set by the ground state of the initial Hamiltonian, and the instantaneous passive energy can thus become negative as $\Delta$ is lowered. The coherent evolution during
stage II produces multiple favorable transfer times, and further
optimization can yield extraction even closer to the ergotropy bound.

\subsection{Discharge into a Load}
\label{sec:load_discharge}

\begin{figure}[t]
    \centering
    \includegraphics[width=.8\linewidth]{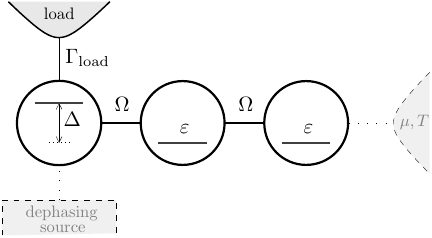}
    \caption{
Discharge protocol by coupling an electronic load. After charging, the
dephasing source and the original charging reservoir are decoupled, and the
detuned QD1 is coupled to a load with tunnel coupling $\Gamma_{\rm load}$.
The load is characterized by temperature $T_{\rm load}$ and chemical
potential $\mu_{\rm load}$. Since the charged state has large population in
the detuned eigenstate $|d\rangle$, coupling the load to QD1 provides a
direct extraction channel.
}
    \label{fig:setup_discharge}
\end{figure}

The unitary protocol above establishes that nearly all of the ergotropy stored
during charging is accessible through coherent gate control. We now consider
a complementary mesoscopic discharge protocol in which the stored energy is
transferred directly to an electronic load, doing work by moving the electron against an electrical potential. After the battery has reached its
charged nonequilibrium steady state, the dephasing source and the charging
reservoir are switched off, and the detuned QD1 is coupled to a separate load reservoir. The load reservoir is characterized by a chemical potential $\mu_{\rm load}$, temperature $T_{\rm load}$, and tunnel coupling $\Gamma_{\rm load}$. Since the internal energy of the charged state is dominated by population in the detuned eigenstate $|d\rangle$, the load is coupled to QD1, as shown in Fig.~\ref{fig:setup_discharge}. The load chemical potential is chosen below the energy of this state,
\begin{equation}
\mu_{\rm load}=E_{\rm d}-4.5k_{\rm B}T_{\rm load},
\end{equation}
so that tunneling mostly occurs from $|d\rangle$ into the load. In the results below, we use $k_{\rm B}T_{\rm load}=\Omega$ and $\Gamma_{\rm load}=0.1\,\Omega$. Later in this section, we examine the dependence of the extracted energy on $\mu_{\rm load}$ and show that this choice lies close to the optimal value.

During discharge, the density matrix evolves according to the same global
master-equation framework introduced in Sec.~\ref{sec:master_equation}, but
with the charging-reservoir and dephasing terms removed:
\begin{equation}
\frac{d\hat\rho}{dt}
=
-\frac{i}{\hbar}[\hat H_0,\hat\rho]
+
\mathcal L_{\rm load}\hat\rho.
\label{eq:discharge_master_equation}
\end{equation}
Since the load is tunnel-coupled to QD1, its coupling to each single-electron
eigenstate $|a\rangle$ is weighted by
$|u_a|^2$. The load Liouvillian is
\begin{equation}
\begin{aligned}
\mathcal L_{\rm load}\hat\rho
=
\frac{1}{\hbar}\sum_a\left(
W_{a0}^{\rm (load)}
\mathcal D[|a\rangle\langle0|]
+
W_{0a}^{\rm (load)}
\mathcal D[|0\rangle\langle a|]
\right)\hat\rho,
\end{aligned}
\label{eq:load_liouvillian}
\end{equation}
where $a\in\{g,e,d\}$. The corresponding extraction and injection rates are,
respectively,
\begin{equation}
\begin{aligned}
W_{0a}^{(\rm load)}
&=
\Gamma_{\rm load}|u_a|^2
[1-f_{\rm load}(E_a)],\\
W_{a0}^{(\rm load)}
&=
\Gamma_{\rm load}|u_a|^2
f_{\rm load}(E_a),
\label{eq:load_rates}
\end{aligned}
\end{equation}
where
$f_{\rm load}(E)=[1+\exp((E-\mu_{\rm load})/k_{\rm B}T_{\rm load})]^{-1}$.
The first process removes an electron from eigenstate $|a\rangle$ and
transfers it to the load, whereas the second injects an electron from the
load into the TQD.

Writing $p_a(t)=\langle a|\hat\rho(t)|a\rangle$ and
$p_0(t)=\langle0|\hat\rho(t)|0\rangle$, the net particle current into the
load is
\begin{equation}
I_{\rm load}(t)
=
\frac{1}{\hbar}\sum_a
\left[
W_{0a}^{(\rm load)}p_a(t)
-
W_{a0}^{(\rm load)}p_0(t)
\right].
\label{eq:load_particle_current}
\end{equation}
The first contribution corresponds to electrons leaving the TQD and entering
the load, while the second describes electrons injected back into the TQD.
Positive $I_{\rm load}$ therefore denotes net electron transfer into the
load. Each electron transferred into the load against the chemical-potential
difference $\mu_{\rm load}-\mu_{\rm source}$ contributes electrochemical work.
The corresponding output power is therefore
\begin{equation}
P_{\rm load}(t)
=
(\mu_{\rm load}-\mu_{\rm source})I_{\rm load}(t),
\label{eq:load_power}
\end{equation}
and the accumulated useful energy delivered to the load is
\begin{equation}
W_{\rm load}(t)
=
\int_0^t P_{\rm load}(t')\,dt'.
\label{eq:integrated_load_output}
\end{equation}
Here $W_{\rm load}$ accounts only for the electrochemical energy stored against
the load bias, rather than the total energy transferred from the TQD, part of
which is dissipated as heat in the reservoir.

\begin{figure}
    \centering
    \includegraphics[width=.9\linewidth]{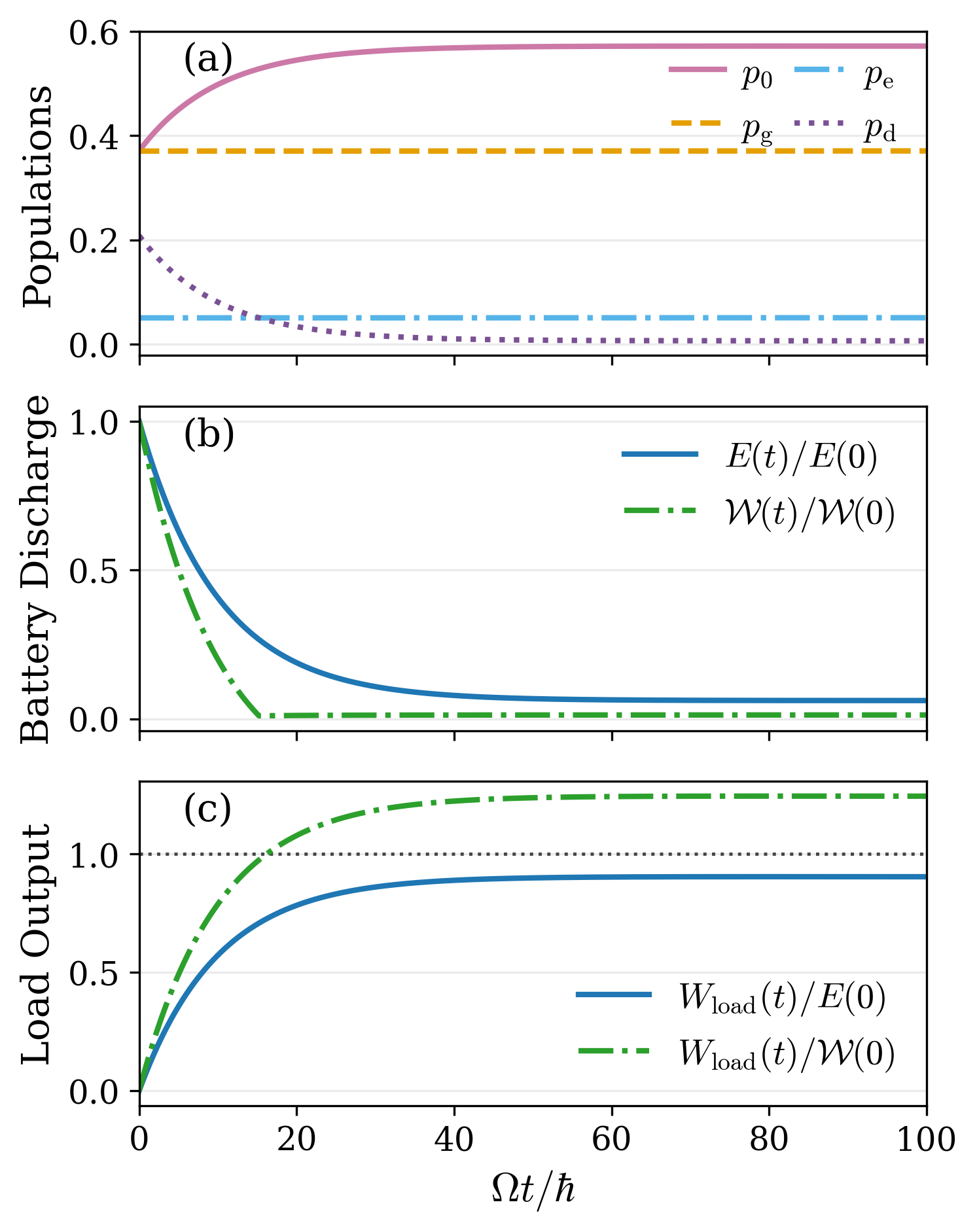}
    \caption{
Load-coupled discharge dynamics. The battery is initialized in the charged
steady state, after which both the dephasing source and the charging reservoir
are switched off, and the detuned dot is coupled to a load reservoir with
$\Gamma_{\rm load}=0.1\Omega$ and
$\mu_{\rm load}=E_{\rm d}-4.5k_{\rm B}T_{\rm load}$.
(a) Eigenstate populations, showing rapid depletion of the detuned state
$p_{\rm d}$.
(b) Normalized internal energy and ergotropy.
(c) Integrated load output, normalized by $E(0)$ and $\mathcal W(0)$.
}
    \label{fig:discharge}
\end{figure}

Figure~\ref{fig:discharge} shows the resulting discharge dynamics.
Panel~(a) gives the eigenstate populations. The population
of the detuned state $p_d$ is rapidly depleted, while the empty-state
population $p_0$ increases. The population of the low-energy state $\ket{e}$
remains nearly unchanged, showing that the load predominantly removes the
high-energy population generated during charging. This selectivity follows
from the strong localization of $\ket{d}$ on QD1 and the much smaller QD1
overlap of the low-energy eigenstates.

This results in the decay of the battery energy and ergotropy, as shown in Fig.~\ref{fig:discharge}(b).
The discharge is much faster than charging because the two processes involve different coupling mechanisms. Charging relies on dephasing-induced transitions into the detuned state, with rates of order $\gamma\Omega^2/(\hbar\Delta^2)$, giving $\tau_{\mathcal W}\propto\hbar\Delta^2/(\gamma\Omega^2)$. During discharge, the load is directly coupled to QD1, where the detuned state is predominantly localized. Its tunneling rate is therefore of order $\Gamma_{\rm load}/\hbar$, giving a characteristic discharge time $\tau_{\rm dis}\sim\hbar/\Gamma_{\rm load}$.
The internal energy saturates at a small nonzero value, reflecting the passive
energy remaining in the final mixed state.

The accumulated useful output $W_{\rm load}(t)$, normalized by both the
initial internal energy $E(0)$ and the initial ergotropy $\mathcal W(0)$, is
plotted in Fig.~\ref{fig:discharge}(c). The normalization by $E(0)$ gives the
fraction of the initially stored energy converted into electrochemical energy
in the load, while normalization by $\mathcal W(0)$ provides a comparison
with the maximum work extractable through cyclic unitary operations. In
contrast to the protocol of Sec.~\ref{sec:unitary_discharge}, the load-coupled
protocol is an open, nonunitary process and is therefore not subject to the ergotropy bound. Particle exchange with the load can also convert
energy belonging to the passive component of the initial state, allowing
$W_{\rm load}$ to exceed the initial ergotropy.

\begin{figure}[t]
    \centering
    \includegraphics[width=.9\linewidth]{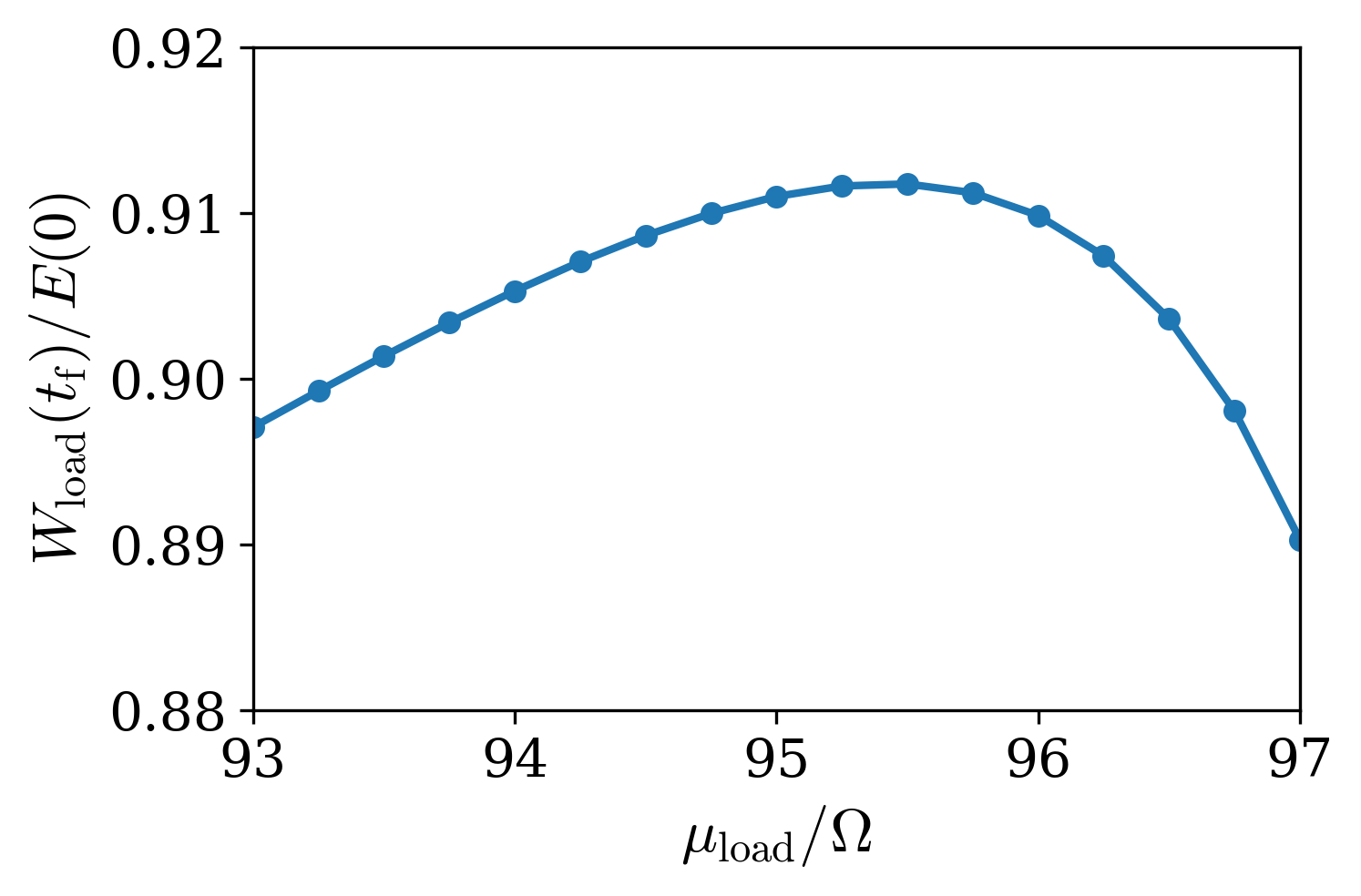}
    \caption{
Dependence of the accumulated useful output on the load chemical potential.
The output $W_{\rm load}(t_{\rm f})$ is evaluated at the end of the discharge
protocol, $\Omega t_{\rm f}/\hbar=100$, with all other parameters as in
Fig.~\ref{fig:discharge}.
}
    \label{fig:load_mu}
\end{figure}

To examine the dependence of the discharge performance on the load bias, we
calculate the accumulated output at the final discharge time,
$W_{\rm load}(t_{\rm f})$, for different values of $\mu_{\rm load}$ with
$\Omega t_{\rm f}/\hbar=100$, as shown in Fig.~\ref{fig:load_mu}. The extracted
energy exhibits a broad maximum around
$\mu_{\rm load}\simeq95.5\,\Omega$. Increasing $\mu_{\rm load}$ initially
increases the electrochemical energy gained per transferred electron, whereas
as $\mu_{\rm load}$ approaches $E_{d}$, tunneling from the detuned state
into the load becomes increasingly suppressed by the Fermi occupation of the
load. The value
$\mu_{\rm load}=E_{d}-4.5k_{\rm B}T_{\rm load}$ used above lies close to
this maximum, with only a negligible reduction in the extracted energy. The
broad maximum also shows that efficient discharge does not require fine
tuning of the load chemical potential.

{An important consideration for repeated operation is whether the state remaining after discharge significantly affects the subsequent charging dynamics. We therefore calculated the charging time (discussed in Sec.~\ref{sec:charging_time}) using the discharged state as the initial condition. The time required to recover $95\%$ of the steady-state ergotropy is $\Omega\tau_{\mathcal W}^{\rm re}/\hbar\simeq 2.62\times10^4$ (for the same parameters as in Fig.~\ref{fig:ergotropy}), compared with $\Omega\tau_{\mathcal W}/\hbar\simeq 2.64\times10^4$ when charging from equilibrium. Thus, recharging after discharge takes about the same time as charging the battery from equilibrium.

The extracted energy ultimately originates from the nonequilibrium dephasing source used during charging. The TQD therefore acts as an intermediate storage element: dephasing prepares an active population in the high-energy state, and the load subsequently extracts this stored energy through a separate discharge channel.

\section{Conclusion}
\label{sec:conclusion}

We have proposed a simple realization of an open quantum battery powered solely by local dephasing. The battery consists of a triple quantum dot coupled to a thermal reservoir, where a strongly detuned end dot acts as the high-energy storage element. Because the local dephasing operator does not commute with the system Hamiltonian, dephasing redistributes population among the energy eigenstates and autonomously drives the system into a highly active nonequilibrium steady state. The large detuning is essential in this architecture, as it creates a high-energy state that is weakly coupled to the reservoir, suppressing relaxation before the stored energy can be extracted.

The battery exhibits a large steady-state ergotropy, with an ergotropic efficiency approaching 1 in the limit $k_{\rm B}T\ll\Omega$. While the final stored energy depends primarily on the detuning of the storage dot, the charging time can be controlled through the dephasing strength, providing an experimentally accessible tuning parameter. Once charged, the dephasing source can be switched off while the stored energy remains trapped in a long-lived charged state because the high-energy state couples only weakly to the reservoir.


We have also demonstrated two complementary approaches to recovering the stored energy. The first uses a cyclic modulation of the detuned-dot gate to extract nearly all the available ergotropy, requiring no additional quantum system or complex control sequence. The second transfers energy directly to an electronic load through particle exchange, converting part of the battery’s passive energy into electrochemical work and thereby exceeding the work accessible through cyclic unitary operations on the battery alone. This output is robust over a range of load chemical potentials. Furthermore, the battery can be recharged after discharge on essentially the same timescale as the initial charging process, supporting its potential for repeated operation.

Our results establish local dephasing as an autonomous thermodynamic resource
for charging open quantum batteries. More broadly, they demonstrate how the
interplay of local dephasing and energy-selective transport can be harnessed
to generate, store, and extract useful energy in experimentally accessible
mesoscopic quantum-dot platforms.

{An important direction for future work is scaling the proposed mechanism to larger energy-storage capacities. The present model operates in the strong Coulomb-blockade regime, where only one electron can occupy the TQD chain at a time. A direct scaling approach would be to operate multiple batteries in parallel, allowing the total stored energy to grow approximately linearly with the number of units. Alternatively, relaxing the single-electron restriction could allow several high-energy states to be occupied simultaneously within the chain. In this case, however, electron-electron interactions would modify the energy spectrum and could affect both the charging mechanism and the extractable work. Exploring such many-electron architectures would be a natural extension of the present work toward practical quantum batteries.}

\section*{Acknowledgments}
A.N.S. and A.N.J. acknowledge support from the John Templeton Foundation Grant ID 63209. R.S. acknowledges funding from the Spanish Ministerio de Ciencia e Innovaci\'on via grant No. PID2024-157821NB-I00, and through the ``Mar\'{i}a de Maeztu'' Programme for Units of Excellence in R{\&}D CEX2023-001316-M.

\vspace{10pt}
\appendix

\section{Charging by Resistor Noise}
\label{app:noise}

In the main text, the charging mechanism is described broadly in terms of a local dephasing channel acting on the detuned barrier dot. One possible microscopic realization of this dephasing source is a nearby resistive element that produces Johnson--Nyquist voltage fluctuations~\cite{clerk2010quantum}. These fluctuations shift the onsite energies of the capacitively coupled barrier dot and thereby generate charge dephasing. In this appendix, we show how this noisy-resistor configuration reduces to the dephasing term used in Sec.~\ref{sec:master_equation}.

We write the fluctuating voltage of the resistive element as $\delta V_{\rm res}(t)$. Since the dots are capacitively coupled to this voltage source, the corresponding interaction Hamiltonian is
\begin{equation}
\hat H_{\rm int}(t)=
\delta V_{\rm res}(t)\hat A_{\rm res},
\qquad
\hat A_{\rm res}=
\sum_{j=1}^{3}
\eta_j \hat n_j .
\label{eq:noise_interaction}
\end{equation}
Here $\hat n_j$ is the occupation operator of dot $j$, and $\eta_j$ is the electrostatic coupling that converts a voltage fluctuation into an energy shift of dot $j$. In a simple lever-arm description, $\eta_j\simeq e\alpha_j$, with $\alpha_j\simeq C_{j{\rm res}}/C_{\Sigma j}$, where $C_{j{\rm res}}$ is the capacitance between dot $j$ and the noisy resistive element, and $C_{\Sigma j}$ is the total capacitance of dot $j$~\cite{kouwenhoven1997electron,vanderwiel2002electron}.

In order for the noisy environment to be represented by the same time-local Lindblad formalism used in the main text, its correlation time must be short compared with the intrinsic timescales of the dot dynamics~\cite{vacchini2024open}. Otherwise, the system evolution would depend on the earlier history of the voltage fluctuations, and the reduced dynamics would require a non-Markovian description. The Markovian limit is therefore obtained when the voltage noise is effectively white over the transition frequencies relevant to the dot system. With the two-sided angular-frequency convention,
\begin{equation}
S_V(\omega)=
\int_{-\infty}^{\infty}
d\tau\,
e^{i\omega\tau}
\langle
\delta V_{\rm res}(\tau)\delta V_{\rm res}(0)
\rangle .
\label{eq:SV_def}
\end{equation}
If $S_V(\omega)\simeq S_V$ is approximately constant over the relevant frequency window, then
\begin{equation}
\langle
\delta V_{\rm res}(t)\delta V_{\rm res}(t')
\rangle=
S_V\delta(t-t') .
\label{eq:white_corr}
\end{equation}
This delta-correlated form shows that the voltage noise is memoryless on the timescales resolved by the quantum-dot dynamics.

For an impedance $Z_{\rm res}(\omega)$, the quantum Johnson--Nyquist spectrum~\cite{johnson_thermal_1928,nyquist_thermal_1928} is
\begin{equation}
S_V(\omega)=
\frac{
2\hbar\omega\,{\rm Re} Z_{\rm res}(\omega)
}{
1-\exp[-\hbar\omega/(k_{\rm B}T_{\rm res})]
},
\label{eq:quantum_JN}
\end{equation}
where $T_{\rm res}$ is the resistor temperature. The white-noise limit follows from two approximations. First, the resistor temperature must be large compared with the relevant transition energies, $k_{\rm B}T_{\rm res}\gg \hbar|\omega|$, so that the quantum spectrum becomes classical. In this limit, $S_V(\omega)\simeq 2k_{\rm B}T_{\rm res}{\rm Re} Z_{\rm res}(\omega)$. Second, the electromagnetic environment must be approximately Ohmic over the same frequency window, ${\rm Re}Z_{\rm res}(\omega)\simeq R_{\rm res}$. Together these assumptions give the frequency-independent spectrum
\begin{equation}
S_V(\omega)\simeq S_V=
2k_{\rm B}T_{\rm res}R_{\rm res}.
\label{eq:white_SV}
\end{equation}

Averaging over the classical white voltage noise gives
\begin{equation}
\frac{d\hat{\rho}}{dt} =
-\frac{i}{\hbar}[\hat H,\hat{\rho}]-
\frac{S_V}{2\hbar^2}
[\hat A_{\rm res},[\hat A_{\rm res},\hat{\rho}]] .
\label{eq:noise_double_commutator}
\end{equation}
Since $\hat A_{\rm res}$ is Hermitian, $\mathcal D[\hat A_{\rm res}]\hat{\rho}=-[\hat A_{\rm res},[\hat A_{\rm res},\hat{\rho}]]/2$, and therefore
\begin{equation}
\frac{d\hat{\rho}}{dt}=
-\frac{i}{\hbar}[\hat H,\hat{\rho}]
+
\frac{S_V}{\hbar^2}
\mathcal D[\hat A_{\rm res}]\hat{\rho} .
\label{eq:noise_lindblad}
\end{equation}
This has the same Lindblad structure as the local dephasing term in Eq.~\eqref{eq:global_master_equation}. Defining
\begin{equation}
\hat L_{\rm res}=
\sqrt{\frac{S_V}{\hbar}}\hat A_{\rm res}
=
\sum_{j=1}^{3}
\eta_j\sqrt{\frac{S_V}{\hbar}}
\hat n_j ,
\label{eq:Lres_local}
\end{equation}
the noisy-resistor contribution can be written as
$\mathcal L_{\rm res}\hat{\rho}
=\hbar^{-1}\mathcal D[\hat L_{\rm res}]\hat{\rho}$,
consistent with the energy-unit convention of the main text. Thus, the effective local dephasing parameters are
\begin{equation}
\gamma_j^{({\rm res})}=
\frac{\eta_j^2 S_V}{\hbar}.
\label{eq:gamma_noise}
\end{equation}
Using Eq.~\eqref{eq:white_SV}, this becomes $\gamma_j^{({\rm res})}=2\eta_j^2 k_{\rm B}T_{\rm res}R_{\rm res}/\hbar$. For $\eta_j=e\alpha_j$, one obtains $\gamma_j^{({\rm res})}=2e^2\alpha_j^2 k_{\rm B}T_{\rm res}R_{\rm res}/\hbar$. For the charging configuration used in the main text, the dominant capacitive coupling is to QD1, so that $\gamma_1^{({\rm res})}=\gamma$ and the remaining dephasing parameters are negligible.

In this realization, the thermodynamic resource is the noisy electromagnetic environment. The resistor supplies voltage fluctuations that dephase the quantum-dot charge states and drive transitions in the energy eigenbasis whenever the dephasing operator does not commute with the TQD Hamiltonian. The white-noise limit used above requires the resistor spectrum to be approximately flat over the relevant transition frequencies. In the present high-detuning regime this requires, in particular, $k_{\rm B}T_{\rm res}\gg\Delta$. Since the electronic reservoir temperature used in the main text satisfies $k_{\rm B}T=\Omega$ with $\Delta=100\Omega$, this corresponds to a resistor temperature much larger than the electronic reservoir temperature. The charging process is therefore powered by an additional hot environment and does not violate the second law.

It is useful to estimate the resistor temperature required to realize the dephasing strengths used in the main text. From Eq.~\eqref{eq:gamma_noise}, the condition $\gamma_j^{({\rm res})}\sim10\Omega$ gives
\begin{equation}
T_{\rm res}
\sim
\frac{5\hbar\Omega}
{e^2\alpha_j^2 k_{\rm B}R_{\rm res}}.
\end{equation}
Taking $R_{\rm res}=R_K=h/e^2$ gives the optimistic mesoscopic high-impedance estimate
\begin{equation}
T_{\rm res}
\sim
\frac{5\Omega}{2\pi\alpha_j^2 k_{\rm B}} .
\end{equation}
For the scale $\Omega=1\,\mu{\rm eV}$ and $\gamma=10\Omega$, this gives $T_{\rm res}\sim92\,{\rm K}$ for $\alpha_j=0.01$, $T_{\rm res}\sim10\,{\rm K}$ for $\alpha_j=0.03$, and $T_{\rm res}\sim0.9\,{\rm K}$ for $\alpha_j=0.1$. These estimates scale inversely with both $R_{\rm res}$ and $\alpha_j^2$. Thus, if the relevant environmental impedance were a conventional $50$~Ohm line rather than a high-impedance mesoscopic environment, the required temperatures would be larger by $R_K/(50\text{\,Ohm})\simeq5\times10^2$. In addition, realizing the classical white-noise limit over the detuning scale requires $k_{\rm B}T_{\rm res}\gg\Delta$. For the parameters used in the main text, $\Delta=100\Omega$, so $\Delta/k_{\rm B}\simeq1.2\,{\rm K}$ when $\Omega=1\,\mu{\rm eV}$. A lever arm of approximately $\alpha_j=0.03$ therefore provides a more moderate temperature estimate while keeping the resistor temperature substantially above the detuning energy scale.

\section{Charging by Continuous Measurement}
\label{app:QPC}

A second microscopic realization of the dephasing source is continuous charge measurement with a nearby quantum point contact (QPC)~\cite{ihn_quantum_2009}. In this case, the QPC current depends weakly on the charge occupation of the nearby dots. Information about the dot charge is therefore continuously acquired by the detector, and the corresponding measurement backaction produces dephasing in the quantum-dot system. In this appendix, we show how the ensemble-averaged dynamics reduce to the same local dephasing channel used in the main text.

Let the QPC be capacitively coupled to the charge configuration of the TQD through the operator
\begin{equation}
\hat A_{\rm QPC}=
\sum_{j=1}^{3}
\lambda_j \hat n_j ,
\label{eq:A_QPC}
\end{equation}
where $\lambda_j$ describes how sensitively the QPC responds to occupation of dot $j$~\cite{singh2026collective}. The resulting dephasing depends on both these coupling coefficients and the detector noise. For the charging configuration considered in the main text, the detector is positioned so that its dominant charge sensitivity is to the detuned QD1, while its coupling to QDs 2 and 3 is negligible.

For a weak continuous measurement, the conditional state of the TQD evolves stochastically because the measurement record is noisy~\cite{andrew_book}. However, when one averages over all possible measurement records, the unconditional density matrix obeys a deterministic master equation. The measurement contribution has the Lindblad form
\begin{equation}
\left.\frac{d\hat\rho}{dt}\right|_{\rm QPC}
=
\mathcal L_{\rm QPC}\hat\rho
=
\frac{1}{\hbar}\mathcal D[\hat L_{\rm QPC}]\hat\rho,
\label{eq:QPC_lindblad}
\end{equation}
with
\begin{equation}
\hat L_{\rm QPC}=
\sum_{j=1}^{3}
\sqrt{\gamma_j^{({\rm QPC})}}\,
\hat n_j .
\label{eq:LQPC}
\end{equation}
Here $\gamma_j^{({\rm QPC})}$ is the measurement-induced dephasing parameter expressed in energy units, consistent with the convention adopted in the main text. The corresponding dephasing rates are proportional to $\gamma_j^{({\rm QPC})}/\hbar$ and are determined by the detector response and shot noise.

For example, in a linearly responding QPC, the measurement rate is proportional to the square of the difference between the detector currents associated with different charge states, divided by the detector noise power $S_I$~\cite{andrew_book}. For an ideal detector, a commonly used expression for the corresponding measurement-induced energy scale is
\begin{equation}
\gamma_j^{({\rm QPC})}
=
\frac{\hbar(I_0-I_j)^2}{4S_I}
=
\frac{eV}{4\pi}
\frac{({\cal T}_0-{\cal T}_j)^2}
{{\cal T}(1-{\cal T})},
\label{eq:QPC_measurement_rate}
\end{equation}
where ${\cal T}_0$ (${\cal T}_j$) is the transmission coefficient of the QPC when QD$j$ is empty (filled), ${\cal T}=({\cal T}_0+{\cal T}_j)/2$, and $V$ is the voltage bias across the QPC. The precise relationship between measurement and dephasing rates depends on the detector operating regime and noise convention. The Markovian description further assumes that the detector noise spectrum is sufficiently broad over the relevant TQD transition frequencies.

In the configuration used in the main text, the detector is sensitive predominantly to QD1. The ensemble-averaged measurement contribution then has the Lindblad form \eqref{eq:QPC_lindblad}
with
\begin{equation}
\hat L_{\rm QPC}
=
\sqrt{\gamma}\,\hat n_1 .
\end{equation}
It suppresses coherences between states with different barrier-dot occupation, while leaving the total particle number unchanged.

The continuous-measurement realization has a different thermodynamic interpretation from the noisy-resistor realization discussed in Appendix~\ref{app:noise}. Here the thermodynamic resource is the biased detector. The QPC is driven out of equilibrium by the applied voltage bias, and the resulting measurement backaction produces the dephasing channel acting on the TQD. The detector record could in principle be used to define conditional trajectories, but the main text only uses the ensemble-averaged dynamics. Therefore, the charging mechanism does not rely on feedback, postselection, or measurement-conditioned work extraction; it relies only on the unconditional dephasing generated by the detector.

\section{Other Dephasing Configurations}
\label{app:deph_config}

In the main text, we focus on dephasing of the detuned QD1. This choice acts directly on the dot that dominates the high-energy storage state $\ket{d}$ and provides the minimal implementation of dephasing-induced charging. Here we compare this configuration with dephasing applied separately to each of the other two dots in the TQD.

For a dephased dot $j$, we use the local operator
\begin{equation} 
\hat L_j=\sqrt{\gamma}\,\hat n_j.
\end{equation}
We compare dephasing of QD1, QD2, and QD3 at the same dephasing strength $\gamma$, together with the evolution in the absence of dephasing.

\begin{figure}
    \centering
    \includegraphics[width=1\linewidth]{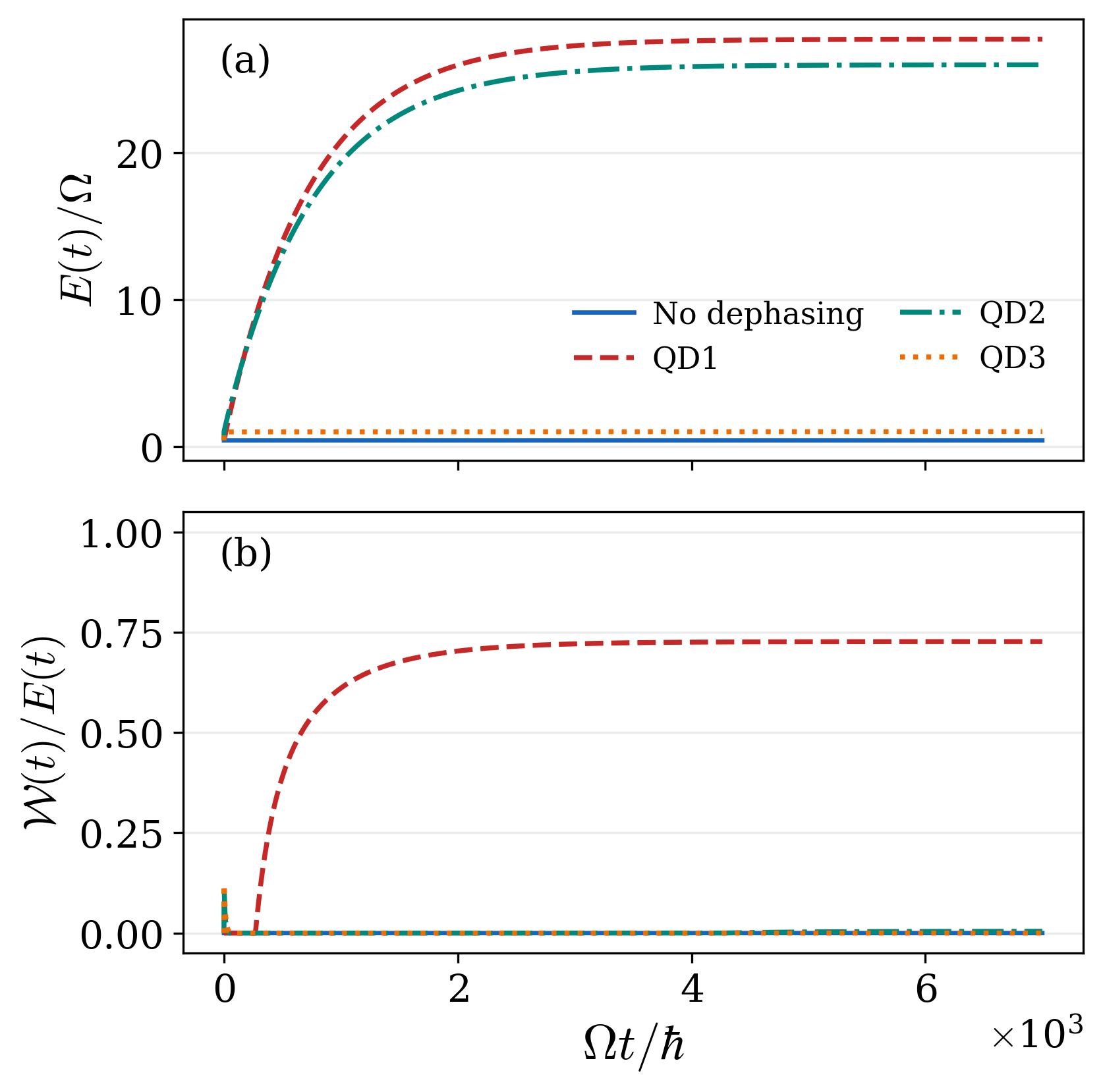}
    \caption{
Charging dynamics for different local dephasing configurations. Dephasing is applied individually to QD1, QD2, or QD3 with the same strength $\gamma$. (a) Internal energy $E(t)/\Omega$. (b) Ergotropic efficiency $\mathcal W(t)/E(t)$. Dephasing of the detuned storage dot QD1 produces both large stored energy and high ergotropic efficiency. Dephasing of QD2 also substantially increases the internal energy but produces little ergotropy, whereas dephasing of the reservoir-coupled QD3 results in negligible charging. The evolution without dephasing is shown for comparison. Parameters are the same as in Fig.~\ref{fig:ergotropy}.
}
    \label{fig:other_configs}
\end{figure}

Figure~\ref{fig:other_configs} shows that the location of the dephasing is crucial not only for the amount of energy deposited in the TQD but also for whether that energy is useful. Dephasing QD1 produces both the largest stored energy and a strongly non-passive steady state, with an ergotropic efficiency approaching the value discussed in the main text. Dephasing QD2 can also inject a substantial amount of energy, but the resulting state is almost passive and therefore has only a small ergotropic fraction. By contrast, dephasing QD3, which is directly connected to the reservoir side of the TQD, produces very little charging and essentially no steady-state ergotropy.

These results demonstrate that energy injection alone is not sufficient for battery charging in the thermodynamic sense. Useful charging requires the dephasing-induced redistribution to create a non-passive population ordering. Acting directly on the detuned storage dot accomplishes both tasks simultaneously: it efficiently populates the high-energy state $\ket{d}$ while the distant reservoir preferentially equilibrates the low-energy sector. This makes QD1 dephasing the natural and minimal choice for the protocol studied in the main text.

\section{Higher-Electron Sector}
\label{app:twoelectron}

The main text uses the strong-Coulomb-blockade description in which the accessible Hilbert space consists of the empty state and the single-electron eigenstates of the triple-dot Hamiltonian. This approximation assumes that higher-charge sectors remain energetically inaccessible through particle injection from the reservoir. However, dephasing can populate the strongly detuned state even when its energy is $\Delta=100\Omega$. Since realistic charging energies can be comparable to the detuning, it is necessary to check whether the same dephasing-induced redistribution could indirectly lead to appreciable occupation of higher-charge sectors. We therefore extend the model to include all two-electron states and explicitly test whether they become populated.

For this purpose, we use a spinful extended-Hubbard Hamiltonian for the
three-dot array, including onsite and distance-dependent interdot Coulomb
interactions~
\cite{burkard_semiconductor_2023,ihn_quantum_2009},
\begin{equation}
\begin{aligned}
\hat H_{\rm ext}
=&
\sum_{i,\sigma}
\epsilon_i
\hat n_{i\sigma} - \Omega
\sum_{i=1}^{2}
\sum_{\sigma}
\left(
\hat c_{i\sigma}^{\dagger}
\hat c_{i+1,\sigma}
+
\hat c_{i+1,\sigma}^{\dagger}
\hat c_{i\sigma}
\right) \\
& + U
\sum_{i=1}^{3}
\hat n_{i\uparrow}
\hat n_{i\downarrow} +
\frac{1}{2}\sum_{i\neq j}
V_{|i-j|}
\hat n_i
\hat n_j ,
\end{aligned}
\label{eq:extended_hubbard}
\end{equation}
where
\begin{equation}
\hat n_i=\hat n_{i\uparrow}+\hat n_{i\downarrow}.
\end{equation}
The barrier dot is detuned by setting
$
\epsilon_1=\Delta,\,
\epsilon_2=\epsilon_3=0.
$
The onsite charging energy is chosen to be the largest interaction scale,
$U=3\Delta$, while the offsite Coulomb interactions are taken to decrease with dot separation according to
\begin{equation}
V_1=\Delta,
\qquad
V_2=0.8\Delta.
\end{equation}
The chosen hierarchy,
$U>V_1>V_2$, reflects the reduction of electrostatic interaction
with dot separation, while the specific ratios are selected as a
conservative test rather than as a fit to a particular device
~\cite{sanchez_longrange_2014,knorzer2022longrange}. Here we use comparatively small interaction energies, of order the detuning, to provide a stringent test of whether multi-electron charging can occur. Larger charging energies would further suppress occupation of higher-particle-number sectors.
These values still represent a strongly interacting regime within the dimensionless model: the mutual charging energies are comparable to the barrier detuning and are much larger than the coherent tunneling, reservoir coupling, and thermal energy scales used in the main text. With $\Delta=100\Omega$, $\Gamma=0.1\Omega$, $k_{\rm B}T=\Omega$, and $\mu=\varepsilon=0$ in the energy convention adopted here, injecting a second electron is energetically costly even though the charging energies are of the same order as the detuned barrier energy.

The many-body Hamiltonian in Eq.~\eqref{eq:extended_hubbard} conserves particle number. We diagonalize it in the truncated Fock space containing all states with $N=0,1,2$. For three spinful dots, this gives
$\binom{6}{0}+\binom{6}{1}+\binom{6}{2}=22$
many-body basis states. This truncation directly tests the first leakage channel beyond the single-electron model: if the $N=2$ sector remains negligible, then higher-charge sectors, which require additional positive charging energies from $U$ and $V_{|i-j|}$, are expected to be even more strongly suppressed. Thus, the $N\le2$ calculation provides a conservative check of the strong-Coulomb-blockade approximation while explicitly allowing all two-electron spin and charge configurations.

The reduced density matrix is evolved in the many-body energy eigenbasis using the same global-master-equation structure as in Sec.~\ref{sec:master_equation}. The reservoir term allows transitions between states whose particle numbers differ by one, with injection and extraction rates determined by the Fermi functions of the reservoirs. The dephasing is included as an exact Lindblad term with Hermitian jump operators proportional to the dot occupations,
\begin{equation}
\hat L_j=\sqrt{\gamma_j}\hat n_j .
\end{equation}
For the configuration used in the main text,
\begin{equation}
\gamma_1=10\Omega,
\qquad
\gamma_2=\gamma_3=0.
\end{equation}
Importantly, dephasing does not itself inject particles into the dot array. Since $\hat n_j$ commutes with the total number operator $\hat N$, the dephasing-induced Lindblad term redistributes population and coherences within a fixed charge sector but does not change $N$. Changes in the total charge can only occur through the reservoir jump processes. Therefore, occupation of the two-electron subspace remains controlled by the reservoir chemical potential and by the charging-energy cost of adding an electron.

\begin{figure}[]
\centering
\includegraphics[width=\linewidth]{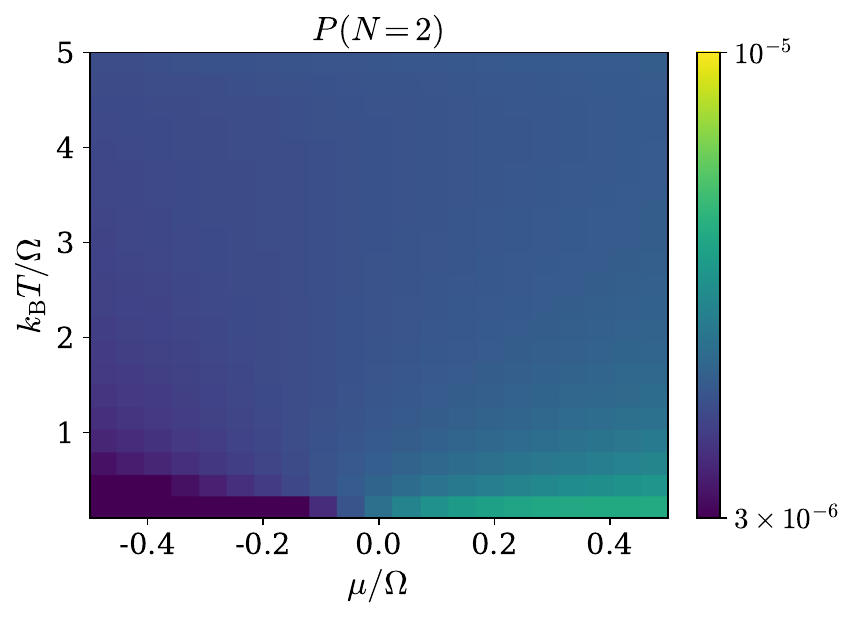}
\caption{
Robustness of the single-electron charging mechanism against two-electron leakage. The calculation includes all $N=0,1,2$ many-body states of the spinful extended-Hubbard Hamiltonian in Eq.~\eqref{eq:extended_hubbard}, with $\Delta=100\Omega$, $U=3\Delta$, $V_1=\Delta$, $V_2=0.8\Delta$, and $\Gamma=0.1\Omega$. The plot shows the steady-state probability of occupying the two-electron sector, $P(N=2)$, over the reservoir parameter window $\mu/\Omega\in[-0.5,0.5]$ and $k_{\rm B}T/\Omega\in[0.1,5]$. The probability remains of order $10^{-6}$–$10^{-5}$, showing that the two-electron sector is strongly suppressed throughout the parameter range considered.
}
\label{fig:twoelectron}
\end{figure}

Figure~\ref{fig:twoelectron} shows the robustness of the two-electron suppression against reservoir parameters. We vary the reservoir chemical potential and temperature over $\mu/\Omega\in[-0.5,0.5]$ and $k_{\rm B}T/\Omega\in[0.1,5]$, while keeping the remaining parameters fixed at the values used in the main text. Throughout this entire parameter window, the two-electron probability remains of order $10^{-6}$–$10^{-5}$, demonstrating that the higher-charge sector is strongly suppressed even when the reservoir temperature is increased to several times the hopping energy.

This result supports the interpretation used in the main text. The dephasing enhances the occupation of the detuned single-electron state and thereby charges the battery, but it does not populate additional electrons by itself. Since the reservoir remains near $\mu=0$ and the two-electron configurations carry large charging penalties, injecting a second electron is energetically unfavorable. The small value of $P(N=2)$ in Fig.~\ref{fig:twoelectron} shows that the dephasing-powered charging mechanism is not a consequence of higher-charge leakage. Rather, it is a redistribution mechanism operating predominantly within the empty and single-electron sectors.

\section{Phonon-Induced Relaxation}
\label{app:phonon_relaxation}

To assess the influence of phonon-assisted relaxation on the battery stability, we supplement the master equation with a phenomenological dissipator describing transitions between the high-energy storage state $\ket{d}$ and the two low-energy eigenstates $\ket{g}$ and $\ket{e}$. The additional contribution is
\begin{equation}
\mathcal L_{\rm ph}\hat\rho
=
\sum_{a=g,e}
\left[
\frac{\kappa_a^\downarrow}{\hbar}
\mathcal D[\ket{a}\bra{d}]\hat\rho
+
\frac{\kappa_a^\uparrow}{\hbar}
\mathcal D[\ket{d}\bra{a}]\hat\rho
\right],
\label{eq:phonon_dissipator}
\end{equation}
where $\mathcal D[\hat L]\hat\rho
=\hat L\hat\rho\hat L^\dagger
-\frac12\{\hat L^\dagger\hat L,\hat\rho\}$.
The downward and upward transition coefficients, expressed in energy units, are
\begin{equation}
\begin{aligned}
\kappa_a^\downarrow
&=
\frac{\kappa_{\rm ph}}{2}
\left[n_{\rm B}(E_d-E_a)+1\right],\\
\kappa_a^\uparrow
&=
\frac{\kappa_{\rm ph}}{2}
n_{\rm B}(E_d-E_a),
\end{aligned}
\label{eq:phonon_rates}
\end{equation}
with
\begin{equation}
n_{\rm B}(E)
=
\frac{1}{e^{E/(k_{\rm B}T_{\rm ph})}-1}.
\end{equation}
Here, $\kappa_{\rm ph}/\hbar$ denotes the total spontaneous emission rate from $\ket{d}$, distributed equally between the two low-energy states. The phonon bath temperature is taken to be $k_{\rm B}T_{\rm ph}=\Omega$, matching the electronic reservoir temperature.

The upward and downward transition rates satisfy detailed balance,
\begin{equation}
\frac{\kappa_a^\uparrow}{\kappa_a^\downarrow}
=
e^{-(E_d-E_a)/(k_{\rm B}T_{\rm ph})}.
\end{equation}
For $\Delta\gg\Omega$ and $k_{\rm B}T_{\rm ph}=\Omega$, phonon absorption into the high-energy state is exponentially suppressed, while spontaneous emission remains possible.

The phonon dissipator is included throughout the charging and storage stages. For each value of $\kappa_{\rm ph}$, the battery is charged under the combined action of the electronic reservoir, local dephasing, and phonon relaxation until its ergotropy reaches $95\%$ of the corresponding steady-state value. The dephasing source is then switched off, while the reservoir and phonon couplings remain unchanged. The resulting evolution is used to evaluate the ergotropy retention shown in Fig.~\ref{fig:stability_phonons}(b).

The phonon-induced transition rates are treated phenomenologically rather than derived from a microscopic electron–phonon Hamiltonian. In a specific device, the corresponding rates depend on the electron--phonon coupling matrix elements, phonon spectral density, and electronic wavefunctions. The calculations therefore characterize the sensitivity of the battery stability to additional inelastic relaxation, rather than predicting the storage lifetime of a particular semiconductor implementation.

\section{Coherent Discharge Protocol}
\label{app:coherentdischarge}

\begin{figure}[t]
    \includegraphics[width=\linewidth]{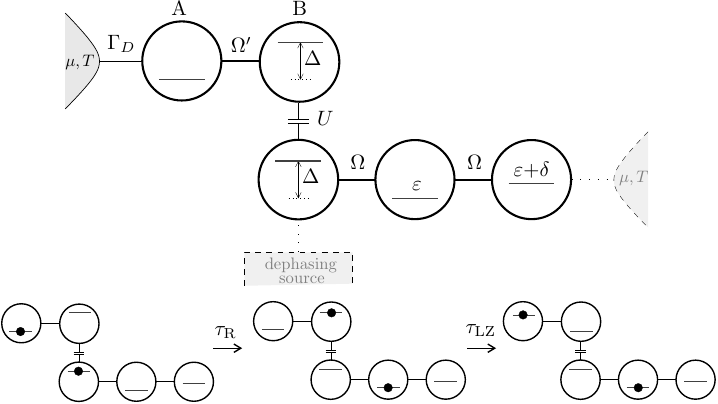}
    \caption{The detuned dot of the TQD battery is capacitively coupled to an auxiliary double quantum dot, enabling coherent transfer of the stored excitation to the auxiliary system.}
    \label{fig:setup_dischargecoh2}
\end{figure}

As an alternative discharge mechanism, we consider a protocol in which QD1 is
capacitively coupled to an auxiliary double quantum dot (DQD), as illustrated
in Fig.~\ref{fig:setup_dischargecoh2}. We denote the two auxiliary dots by $A$
and $B$, with interdot hopping amplitude $\Omega'$. The state basis is augmented by including the occupation of the two subarrays: $|Aj\rangle$ and $|Bj\rangle$ if the electrons occupy quantum dots $A$ or $B$ in the auxiliary array, and $j$ in the battery. The capacitive coupling between QD1 and $B$ ensures that simultaneous occupation of these dots carries an additional
interaction energy $U$. The auxiliary levels are detuned according to
$E_A-E_B=E_2-E_1=-\Delta$, which we parametrize as
$E_A=\varepsilon'$ and $E_B=\varepsilon'+\Delta$.

The auxiliary DQD remains empty during charging of the TQD. Once the battery
has been charged, an electron is loaded into dot $A$ from an attached
reservoir, preparing the initial state $\ket{A1}$. 
At this stage it is convenient to detune the energies of QDs 2 and 3, $E_3-E_2=\delta$, so they become effectively disconnected from the dynamics.
Energy stored in QD1 can
then be coherently transferred to the auxiliary DQD through the cotunneling
process
\begin{equation}
    \ket{A1}\rightarrow\ket{X}\rightarrow\ket{B2},
\end{equation} 
The intermediate configurations $X=A2$ or $X=B1$ are off resonant and therefore only virtually
occupied. The resulting effective coupling between the initial and final
configurations is
\begin{equation}
    \hat{H}_{\rm eff}
    =-\frac{\Omega\Omega' U}{\Delta(\Delta+U)}
   \ket{A1}\bra{B2}
    +{\rm h.c.}
\end{equation}
Thus, the effective coupling, $g_{\rm eff}
=\frac{\Omega\Omega' U}{\Delta(\Delta+U)}$, enables coherent Rabi oscillations. After half a Rabi period, $t_g=\pi\hbar/2g_{\rm eff}$, the
electron initially occupying QD1 is transferred to QD2, while the
electron in the auxiliary DQD is promoted to the higher energy state in $A$. The energy
released by the TQD is thereby transferred to the auxiliary system. 

At this time $t_g$, the two systems must be brought out of
resonance to prevent the reverse process. This can be achieved by sweeping
the auxiliary levels through a Landau--Zener transition, transferring the
electron between the auxiliary dots, followed by a rapid change of the energy of the empty
dot $B$. The final level configuration is chosen such that
$E_A-E_B\neq E_2-E_1$, this way suppressing further resonant cotunneling between the
TQD and the auxiliary DQD.  The energies of dots 2 and 3 are then restored to $\varepsilon$. The protocol therefore provides a concrete route
for coherently transferring the energy stored in the barrier dot to an auxiliary
quantum system while leaving the TQD in a lower-energy configuration.

\bibliography{references}
\end{document}